\documentclass{aa}

\usepackage{epsfig}
\usepackage{graphicx}
\usepackage{times}
\usepackage{color}
\usepackage{ulem}
\usepackage{hyperref}
\usepackage{url}

\usepackage{xcolor}
\hypersetup{
colorlinks,
linkcolor={red!80!black},
citecolor={blue!80!black},
urlcolor={blue!80!black}
}
\def\cm3{cm$^{-3}$}
\def\kms{km~s$^{-1}$}

\def\lsun{L$_{\odot}$}
\def\rsun{R$_{\odot}$}

\def\msun{M$_{\odot}$}
\def\zsun{Z$_{\odot}$}

\def\one{\ts {\,\sc i}}
\def\two{\ts {\,\sc ii}}
\def\three{\ts {\,\sc iii}}
\def\four{\ts {\,\sc iv}}
\def\five{\ts {\sc v}}

\def\beq{\begin{equation}}
\def\eeq{\end{equation}}

\def\lesssim{\mathrel{\hbox{\rlap{\hbox{\lower4pt\hbox{$\sim$}}}\hbox{$<$}}}}
\def\gtrsim{\mathrel{\hbox{\rlap{\hbox{\lower4pt\hbox{$\sim$}}}\hbox{$>$}}}}

\def\lesssim{\mathrel{\hbox{\rlap{\hbox{\lower4pt\hbox{$\sim$}}}\hbox{$<$}}}}
\def\gtrsim{\mathrel{\hbox{\rlap{\hbox{\lower4pt\hbox{$\sim$}}}\hbox{$>$}}}}

\def\one{{\,\sc i}}
\def\two{{\,\sc ii}}
\def\three{{\,\sc iii}}
\def\four{{\,\sc iv}}
\def\five{{\sc v}}

\def\v1d{{\sc v1d}}

\def\mesa{{\sc mesa}}
\def\cmfgen{{\sc cmfgen}}

\newcommand{\iso}[2]{\ensuremath{^{#1}\rm{#2}}}

\def\aj{AJ}
\def\pasp{PASP}

\def\apj{ApJ}
\def\apjs{ApJS}
\def\apjl{ApJL}
\def\aap{A\&A}
\def\araa{ARA\&A}

\def\mnras{MNRAS}
\def\nat{Nature}

\def\nifs{\iso{56}Ni}
\def\cofs{\iso{56}Co}

\begin{document}

\title{Supernovae from blue supergiant progenitors: What a mess!}

\titlerunning{Radiative-transfer models for Type II-pec SNe}

\author{Luc Dessart\inst{\ref{inst1}}
  \and
  D. John Hillier\inst{\ref{inst2}}
  }

\institute{Unidad Mixta Internacional Franco-Chilena de Astronom\'ia
(CNRS UMI 3386), Departamento de Astronom\'ia, Universidad de Chile,
Camino El Observatorio 1515, Las Condes, Santiago, Chile\label{inst1}
    \and
    Department of Physics and Astronomy \& Pittsburgh Particle Physics,
    Astrophysics, and Cosmology Center (PITT PACC),  University of Pittsburgh,
    3941 O'Hara Street, Pittsburgh, PA 15260, USA.\label{inst2}
  }

  \date{Accepted . Received }

\abstract{Supernova (SN) 1987A was classified as a peculiar Type II SN because of its long rising light curve and the persistent presence of H\one\ lines in optical spectra. It was subsequently realized that its progenitor was a blue supergiant (BSG), rather than a red supergiant (RSG) as for normal, Type II-P, SNe. Since then, the number of Type II-pec SNe has grown, revealing a rich diversity in photometric and spectroscopic properties. In this study, using a single 15\,\msun\ low-metallicity progenitor that dies as a BSG, we have generated explosions  with a range of energies and \nifs\ masses. We then performed the radiative transfer modeling with \cmfgen, from 1\,d until 300\,d after explosion for all ejecta. Our models yield light curves that rise to optical maximum in about 100\,d, with a similar brightening rate, and with a peak absolute $V$-band magnitude spanning $-14$ to $-16.5$\,mag. All models follow a similar color evolution, entering the recombination phase within a few days of explosion, and reddening further until the nebular phase. Their spectral evolution is analogous, mostly differing in line width. With this model set, we study the Type II-pec SNe 1987A, 2000cb, 2006V, 2006au, 2009E, and 2009mw. The photometric and spectroscopic diversity of observed SNe II-pec  suggests that there is no prototype for this class. All these SNe brighten to maximum faster than our limited set of models, except perhaps SN\,2009mw. The spectral evolution of SN\,1987A conflicts with other observations in this set and conflicts with model predictions from 20\,d until maximum: H$\alpha$ narrows and weakens while Ba\two\ lines strengthen faster than expected, which we interpret as signatures of clumping. SN\,2000cb rises to maximum in only 20\,d and shows weak Ba\two\ lines. Its spectral evolution (color, line width and strength) is well matched by an energetic ejecta but the light curve may require strong asymmetry. The persistent blue color, narrow lines, and weak H$\alpha$ absorption, seen in SN\,2006V conflicts with expectations for a BSG explosion powered by \nifs\ and may require an alternative power source. In contrast with theoretical expectations, observed spectra reveal a diverse behavior for lines like Ba\two\,6142\,\AA, Na\one\,D, and H$\alpha$. In addition to diversity arising from different BSG progenitors, we surmise that their ejecta are asymmetric, clumped, and, in some cases, not solely powered by \nifs\ decay.
}


\maketitle
\label{firstpage}


 \section{Introduction}
\label{sect_intro}

Although the majority of Type II supernovae (SNe) exhibit a declining
bolometric light curve after shock breakout, compatible with
the explosion of a red-supergiant (RSG) star, a few per cent of Type II SNe
exhibit instead a long rise to optical (and generally bolometric) maximum
\citep{woosley_87a_88,kleiser_2pec_11,pasto_9e_2pec_12,taddia_2pec_12,taddia_2pec_16}.
Owing to this light curve peculiarity, these SNe are classified as Type II-peculiar.
While some SNe with a long-rising optical or bolometric light curve can be very luminous
at maximum (see, e.g., \citealt{terreran_slsn2_17}), numerous SNe II-pec reach a modest
peak $V$-band brightness, even when characterized by optical spectra with broad
(Doppler-broadened) lines.

For SN\,1987A, the founding member of the Type II-pec SN class,
the H$\alpha$ line absorption at one day is maximum at about $-18000$\,\kms\
from line center (\citealt{phillips_87A_88}; this is amongst the largest values ever recorded
for a Type II SN), but the SN peaks about 80\,d after
explosion with a maximum $V$-band brightness of only $-16$\,mag
(\citealt{catchpole_87A_87}; \citealt{hamuy_87A_88}; this is fainter than for typical Type II SNe).
The basic characteristic that distinguishes SN\,1987A and Type II-peculiar SNe
from Type II-P SNe is their smaller progenitor radii \citep{woosley_87a_88,woosley_87A_late_88,shigeyama_nomoto_90,utrobin_87A_93,blinnikov_87A_00, utrobin_etal_15,taddia_2pec_16}. These works also emphasize that the morphology of SN\,1987A and of Type II-pec SN light curves generally requires mixing of \nifs\ into the outer progenitor envelope as well as mixing
of envelope material (rich in H and He) deep  into the progenitor He core.
Such mixing is expected theoretically, inferred from nebular phase spectral modeling,
and seen in multidimensional  hydrodynamical simulations of blue-supergiant (BSG)
star explosions
\citep{ebisuzaki_87A_89,fryxell_mueller_arnett_91,liu_dalgarno_92,li_87A_93,KF98a,kifonidis_00,kifonidis_03,wongwathanarat_13_3d,wongwathanarat_15_3d}.
While these works present compelling evidence that we can reproduce the observations of SNe II-pec,
the modeling has been limited to the photometry, and sometimes only to the bolometric light curve.
Radiation hydrodynamical codes employ a variety of techniques for radiation transport,
usually treating the gas and the radiation as two separate fluids (see, e.g., discussion in
\citealt{blinnikov_87A_00}). The gas is however always assumed to be in local
thermodynamic equilibrium (LTE) and in steady state.

Spectroscopic modeling of SN\,1987A has been done extensively,
with successes and problems, using a steady-state non-LTE approach.
\citet{hoeflich_87A_88} finds evidence for departures from LTE as well as evidence for mixing.
\citet{eastman_87A_89} note a problem with the reproduction of He\one\ lines.
\citet{schmutz_87A_90} discuss the difficulty of matching both the absorption
and emission parts of Balmer lines, alluding to the presence of clumping.
\citet{mitchell_56ni_87A_01} discuss the importance of \nifs\ mixing
for reproducing Balmer lines at early times, as early as four days after explosion.
\citet{mazzali_87A_92} suggest the need for an overabundance of Ba to reproduce
the Ba\two\ lines in the optical at $20-30$ days after explosion.
Nebular phase spectra also suggest strong macroscopic mixing and a clumpy ejecta
\citep{fransson_chevalier_89, spyromilio_87a_90, li_87A_93, jerkstand_87a_11, jerkstrand_04et_12}.

More recently, simulations have shown that time-dependent effects in the non-LTE rate
equations have a critical impact on the strength of Balmer lines and Ba\two\ lines \citep{UC05}.
Because of its impact on the ionization, time dependence
has been found to impact the formation of all lines and the whole spectrum \citep{D08_time}.
With time dependence, the regions at and above the photosphere are predicted
to be more ionized than when steady state is assumed.
The assumption of steady state is therefore a shortcoming of older radiative transfer
calculations, whose consequences on the conclusions of these works is not known.
For example, \citet{schmutz_87A_90} suggest that clumping may be at the origin
of their underestimate of the H$\alpha$ emission, but this may be attributed to
the neglect of time-dependent terms in the non-LTE rate equations.
\citet{d18_fcl} find that clumping tends to reduce the H$\alpha$ line strength,
which is the opposite effect proposed by \citet{schmutz_87A_90}. The dominant
effect of clumping seen in the simulations of \citet{d18_fcl} is the reduction
of the ionization (and therefore the density of free electrons) at and above
the photosphere, and the reduction of the H$\alpha$ emission flux.
\cite{mitchell_56ni_87A_01} argue for the need of \nifs\ mixing to
reproduce Balmer lines at early times but strong Balmer lines can
be explained exclusively from a time-dependent effect \citep{UC05,D08_time}.
These conflicting conclusions about the spectral properties of SN\,1987A
suggest that the consensus about this SN is at best superficial. It is
not clear whether these spectral peculiarities (and our interpretation of their
origin) impact the inferences from
light curve modeling or alter our understanding of the progenitor and explosion physics.
Right now,  SN spectral properties are largely ignored (essentially limited to
estimating the expansion rate and classifying the SN) when inferring the SN properties.
As we emphasize in this paper, the differences in spectral properties for events
sharing the same classification as Type II-pec reveal a rich diversity
of explosion properties, perhaps even of power sources. So, there is more to
Type II-pec SN progenitors than just a reduced radius.

The advantage of performing simulations of the bolometric light curve using
radiation hydrodynamics, or of the spectra using steady-state
radiative transfer, is the speed at which one can generate results and iterate until
a good match is obtained. This advantage is however offset by the lower
level of consistency of the approach since the successful model only reproduces
a fraction of the constraints. In \citet{DH10} and \citet{li_etal_12_nonte}, we
presented non-LTE time-dependent simulations for one ejecta model and compared
to the observations of SN\,1987A. With our approach, we could reproduce H\one\
and He\one\ lines at early times, even in the absence of non-thermal processes.
At the recombination epoch, the time dependent treatment captured the ionization freeze-out
and allowed us to reproduce H\one\ lines. At late times, non-thermal processes
(combined with strong inward mixing of H) are essential to retain a strong H$\alpha$
line. But our studies were limited to one ejecta model. They were also started at 0.3\,d when
the ejecta was not in homologous expansion. To accommodate the assumption of homology
in \cmfgen, we had to force it, which affected the initial structure of our simulations.
Overall, these simulations yielded significant departures from the observations of SN\,1987A,
although qualitatively, its basic photometric and spectroscopic properties were satisfactorily
reproduced in our ab-initio approach.

\begin{table}
\caption{
Characteristics of our selected sample of Type II-pec SNe  (see discussion
and references in Section~\ref{sect_obs}).
\label{tab_obs}
}
\begin{center}
\begin{tabular}{
l@{\hspace{2mm}}c@{\hspace{2mm}}c@{\hspace{2mm}}
c@{\hspace{2mm}}c@{\hspace{2mm}}c@{\hspace{2mm}}
c@{\hspace{2mm}}c@{\hspace{2mm}}c@{\hspace{2mm}}
}
\hline
        SN       &     $D$      &  $\mu$   & $t_{\rm expl}$   &    $E(B-V)$    &    $z$\\
                     &     [Mpc]  &     [mag]  &     MJD [d]           &     [mag]        &        \\
\hline
   SN\,1987A       &   0.05            & 18.50                &  46849.82       &      0.150    &   8.83(-4)  \\
   SN\,2000cb     &  30.0             &  32.39              &   51657.00        &       0.114   &   6.40(-3)     \\
   SN\,2006V      & 72.7               & 34.31                 &    53748.00       &      0.029 &   1.58(-2)      \\
   SN\,2006au     &  46.2             &  33.32                &   53794.00        &      0.312  &   9.80(-3)    \\
   SN\,2009E       &   29.97           & 32.38                &     54832.50      &      0.040  &   7.20(-3)           \\
   SN\,2009mw   &   50.95          &  33.54               &     55174.50       &      0.054     &   1.43(-2)          \\
\hline
\end{tabular}
\end{center}
\end{table}

Here we revisit our previous modeling of BSG star explosions. We present simulations based on one BSG progenitor model whose characteristics are close to those proposed for the prototypical Type II-pec SN\,1987A (in particular, its progenitor radius of about 50\,\rsun; \citealt{sn1987A_rev_90}). However, this model is now used to generate a wide range of ejecta models characterized by different explosion energies, \nifs\ mass, H to He abundance ratio, or chemical mixing. We confront our results to SN\,1987A as well as other Type II-pec SNe observed since, including 2000cb, 2006V, 2006au, 2009E, and 2009mw \citep{kleiser_2pec_11,pasto_9e_2pec_12,taddia_2pec_12,taddia_2pec_16, takats_09mw_16}. This sample of SNe II-pec allows us to step back from SN\,1987A and consider the SN II-pec class at large. Our simulations, based on smooth and spherically-symmetric ejecta, allow us to gauge how clumpy and asymmetric each SN ejecta may be, as well as discuss the power source at the origin of the SN luminosity.

In the next section, we present the set of observations used in this study.
In Section~\ref{sect_method}, we present our numerical approach, describing
how the progenitor model was produced, how the explosion was generated, and
how we compute the radiative properties of the resulting ejecta.
We then present the results from our simulations, starting with the UVOIR light curves
(Section~\ref{sect_luvoir}), followed by the $V$-band and optical color evolution (Section~\ref{sect_photometric}),
and the spectral properties (Section~\ref{sect_spec}).
In Section~\ref{sect_phot_struct}, we comment on the structure of the photospheric layers,
and in particular on the offset between the location of the photosphere and the location
of the recombination front in our Type II-pec SN ejecta.
We then compare our model results with the observed photometric and spectroscopic properties
of Type II-pec SNe (Section~\ref{sect_spec_obs}).
Finally, we present our conclusions in Section~\ref{sect_conc}.

\section{Observations}
\label{sect_obs}

   To compare our models to observations, we use photometric and
   spectroscopic data for a sample of Type II SNe having an optical or bolometric
   light curve with a long rise to maximum. In practice, we selected
   SNe 1987A, 2000cb, 2006V, 2006au, 2009E, and 2009mw (this is not meant to
   be an exhaustive list).  Below, we summarize the basic characteristics of each SN
   (see also Table~\ref{tab_obs}) and the source of the observational data used.
   The data was downloaded from the SN catalog \citep{sn_catalog} at
   \href{https://sne.space}{https://sne.space}
   and from {\sc wiserep}  \citep{wiserep} at
   \href{https://wiserep.weizmann.ac.il}{https://wiserep.weizmann.ac.il}.

\subsection{SN\,1987A}

   For SN\,1987A, we used spectra from \citet{phillips_87A_88} for the first 130\,d and from
   \citet{phillips_87A_90} for later times.
   For the inferred bolometric light curve, we used both the results of
   \citet{hamuy_87A_88} and \citet{catchpole_87A_87} (which differ, in part,
   from the different values of the adopted reddening, i.e., 0.15 and 0.2\,mag).
   For the photometry, we use the data from these papers.
   We adopted the time of the neutrino burst at MJD\,46849.82
   as the time of explosion for SN\,1987A \citep{hirata_87A_87,bionta_87A_87}.
   We used a recession velocity of 286.5\,\kms\ (redshift of 0.00096),
   from \citet{meaburn_87A_95}.
   We adopted  a reddening $E(B-V)=0.15$\,mag and
   a distance modulus of 18.5\,mag (equivalent to a distance of 50\,kpc).

\subsection{SN\,2000cb}

Following \citet{kleiser_2pec_11}, we adopted a
distance of 30.0\,Mpc, a reddening $E(B-V)$ of 0.114\,mag, and a redshift of 0.0064.
We adopt the explosion date MJD\,51657.0 (i.e., one day later than the value adopted by \citealt{kleiser_2pec_11}) since it yields a better match to the early time
spectra and optical color of model a5 (Section~\ref{sect_spec_00cb}).

\subsection{SN2006V}

For SN\,2006V, we use the photometric and spectroscopic data from \citet{taddia_2pec_12}. We adopted a distance of  72.70\,Mpc, a redshift of  0.0158, a reddening $E(B-V)=0.029$\,mag, and an explosion date of  MJD\,53748.0 \citep{taddia_2pec_12}. SN\,2006V shows a number of peculiarities, and most notably its persistent blue optical color and the anomalously weak H$\alpha$ absorption during the first two months after explosion, combined with the absence of O\one\,7774\,\AA.

\subsection{SN\,2006au}
\label{sect_obs_06au}

For SN\,2006au, we use the photometric and spectroscopic data from \citet{taddia_2pec_12}.
We adopted a distance of  46.20\,Mpc,
a redshift of 0.0098, a reddening $E(B-V)=0.312$\,mag, and
the explosion epoch of  MJD\,53794.0 from \citet{taddia_2pec_12}.

This high reddening implies a spectrum with a blue color at the recombination
epoch, just like for SN\,2006V, and in contrast with SN1987A \citep{taddia_2pec_12}.
Physically, a very low, sub-LMC, metallicity (together with a very weak mixing of core
metals into the H-rich layers) might explain such a blue color (see the models of \citealt{d13_sn2p},
and the recent observations of SN\,2015bs for which a very low metallicity of 0.1\,\zsun\
is inferred; \citealt{anderson_15bs}).
SNe 2006V and 2006au are both located far out in the spiral arms of their host galaxy,
so a very low metallicity is possible for both.
Another possibility is that the reddening toward SN\,2006au is in fact negligible.
In this case, the intrinsic brightness and the optical colors of SN\,2006au would be comparable
to those of SN\,1987A (see Section~\ref{sect_spec_06au} for discussion).

\begin{figure}
   \includegraphics[width=\hsize]{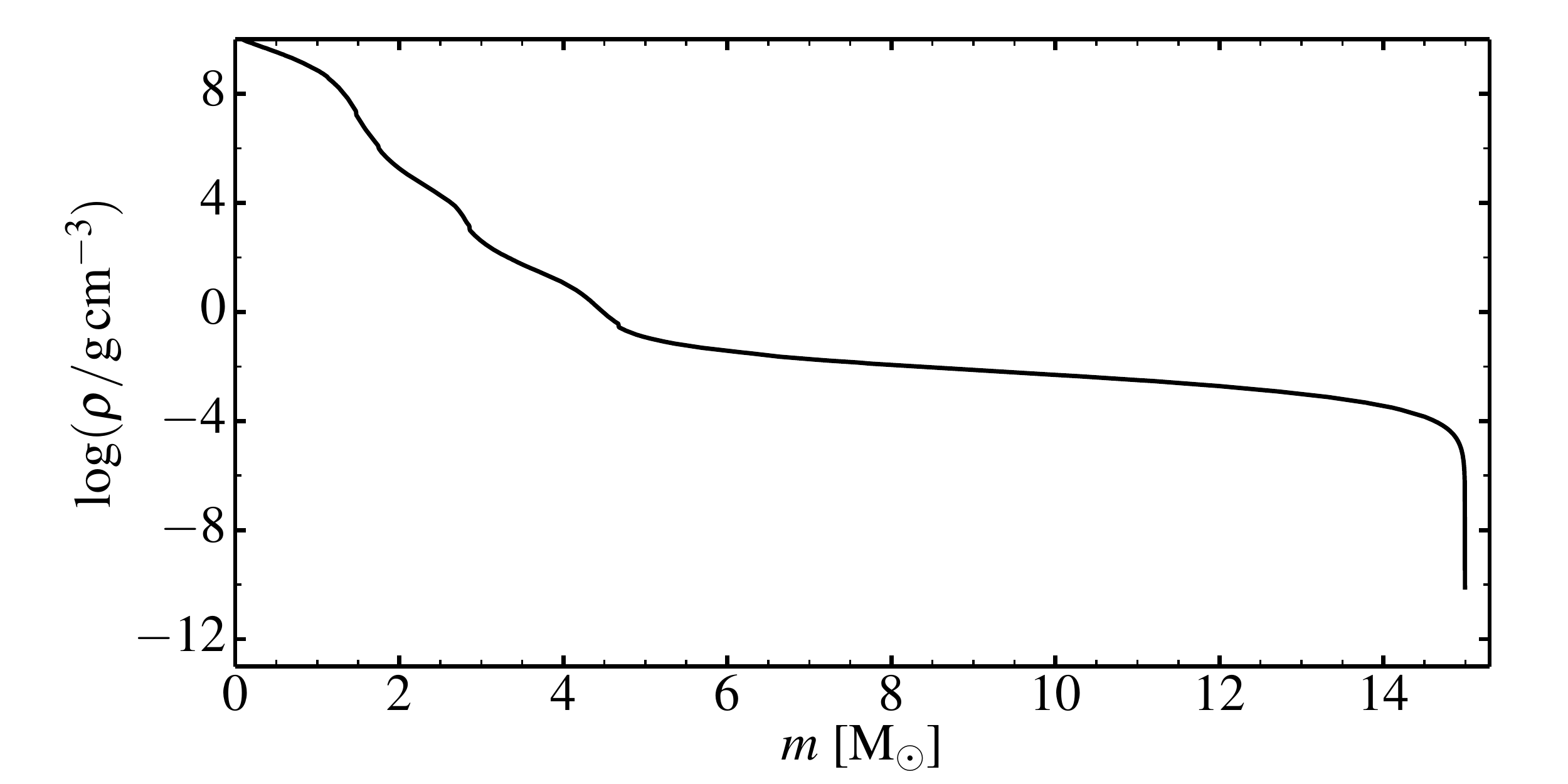}
   \includegraphics[width=\hsize]{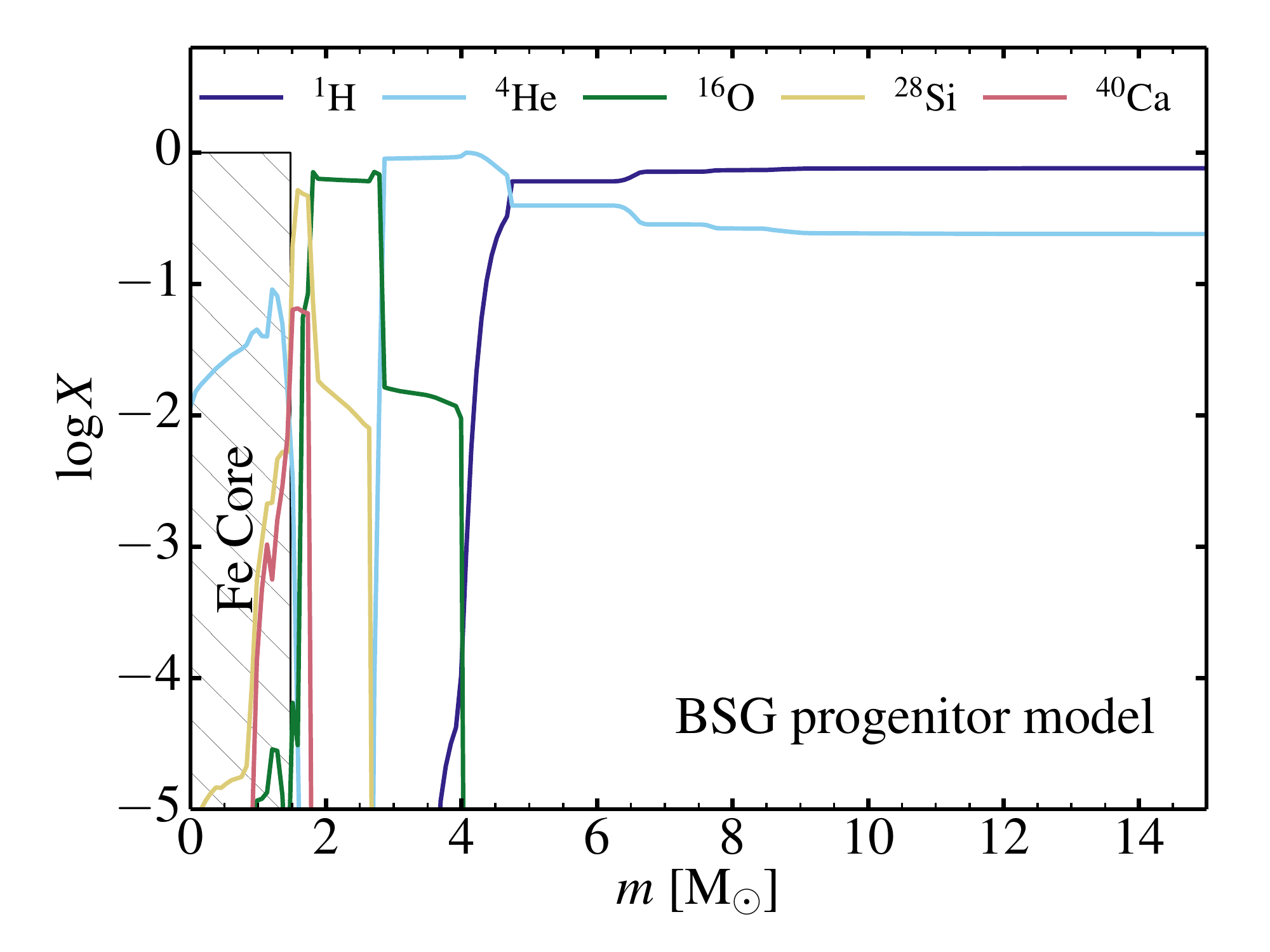}
\caption{
Density (top) and composition (bottom) profiles of our 15\,\msun\ model
at the onset of core collapse.
The model corresponds to a BSG star with a surface radius of 44.8\,\rsun.
The adopted environment metallicity is 10$^{-7}$.
The bulk of the composition is H and He rich, and the metal-rich part is
limited to the He-core below a lagrangian mass of 4\,\msun.
(See Section~\ref{sect_prog} for discussion).
 \label{fig_prog}
}
\end{figure}

\subsection{SN\,2009E}

Following \citet{pasto_9e_2pec_12}, we adopted a distance to SN\,2009E of
29.97\,Mpc, a reddening $E(B-V)=0.04$\,mag, a recession velocity of
2158\,\kms\ (redshift of 0.00720), and the explosion date MJD\,54832.5.
Although it depends on the adopted reddening, the offset in optical brightness
with SN\,1987A is small.

\subsection{SN\,2009mw}

We use the photometric and spectroscopic data for SN\,2009mw of \citet{takats_09mw_16}.
The optical light curve of SN\,2009mw is similar to that of SNe 1987A and 2009E
\citep{takats_09mw_16}.
The spectroscopic data  may have some calibration issues.
The first spectrum has an usual slope long-ward of 8000\,\AA, while the
second epoch has a bluer color than either the first or the third spectra.
In the present comparison we have ignored the second spectrum.
Following \citet{takats_09mw_16}, we adopted a distance of 50.95\,Mpc,
a reddening $E(B-V)$ of 0.054\,mag, a redshift of 0.0143, and an explosion
date of MJD\,55174.5. The explosion date is uncertain.

\section{Initial conditions and methodology}
\label{sect_method}

Our model set is limited to one progenitor star, hence to a single progenitor mass.
We use a 15\,\msun\ progenitor because this represents a standard mass for a massive
star. The progenitor radius of 50\,\rsun\ is typical for a BSG star, which is thought to
be the defining characteristic of SNe II-pec compared to SNe II-P or II-L. We adopt
an ejecta metallicity typical of the LMC since it is thought to be representative of SNe II-pec
\citep{taddia_2pec_16}. We enforce a strong mixing since this property has been inferred by all
light curve calculations of SN\,1987A (see, e.g., \citealt{blinnikov_87A_00}). These
results allow us to narrow the vast parameter space one would ideally like to cover,
which includes a range of progenitor radii, masses, metallicities, explosion energies,
\nifs\ mass, \nifs\ mixing. Covering all permutations would be a major task with any code.

\subsection{The progenitor model}
\label{sect_prog}

Our study is based on a non-rotating zero-age main sequence star of 15\,\msun.
This initial model is evolved with \mesa\ \citep{mesa1,mesa2,mesa3} at a metallicity of 10$^{-7}$.
It reaches core collapse after 12.88\,Myr (the \mesa\ simulation
is stopped when the maximum infall velocity reaches 1000\,\kms).
At that time,  the surface luminosity is 57400\,\lsun, the surface radius is 44.8\,\rsun,
the effective temperature is 13350\,K, the total mass is 14.99\,\msun,
the CO core mass is 2.83\,\msun, the He core mass is 4.07\,\msun,
and the mass of the H-rich envelope is 10.92\,\msun.
Hence, with this very low metallicity the model reaches core collapse as a BSG star
rather than a RSG star without any tinkering of convection parameters or the introduction
of rotation etc.
We show some properties of this model at the onset of core collapse in Fig.~\ref{fig_prog}.

\subsection{Explosion models}
\label{sect_expl}

The explosion is simulated with \v1d\ \citep{livne_93,DLW10a, DLW10b} by setting
up a thermal bomb at the Lagrangian mass cut of 1.55\,\msun.
The binding energy of the overlying envelope is $4 \times 10^{50}$\,erg.
Energy is deposited at a depth-independent rate in the inner 0.05\,\msun\ for 0.5\,s.
This rate is chosen to produce a set of ejecta with a kinetic energy at infinity
of 0.4, 0.8, 1.2, and $2.4 \times 10^{51}$\,erg. In this order, this set corresponds
to model names a2, a3, a4, and a5 (model a1 with half the energy of model a2 exhibits
strong fallback and is excluded from the study). By varying
the explosion energy, we mimic the effect of varying the ejecta (or the progenitor) mass.

In this study, we want to have a \iso{56}Ni mass that increases with explosion energy
and is in rough correspondence to observed values (e.g., for SN\,1987A; \citealt{sn1987A_rev_90}).
In simulations, the \iso{56}Ni mass is sensitive to the explosion trigger (energy,
power, mass cut location) so it is hard to control.
Hence, at 5\,s after the explosion trigger in the \v1d\ simulations, we reset
the \iso{56}Ni mass to be 0.02 (model a2), 0.04 (a3), 0.08 (a4), and 0.16\,\msun\ (a5).
When rescaling the \iso{56}Ni mass fraction, we renormalize all other mass fractions
at each depth so that the sum of mass fractions is unity.
In model a4ni (a3ni), the \iso{56}Ni mass fraction is scaled by a factor of three (two) compared
to model a4 (a3).
In model a4he, we scale by a factor of 0.658 the hydrogen mass fraction in the \mesa\ model
at the time of collapse,  while the difference in total mass fraction at each depth
is transferred to helium to keep the normalization to unity.
This scaled \mesa\ model is then exploded the same way as model a4.

In each model, we imposed a mixing of all species in mass space.
At 5\,s after the explosion trigger, we stepped through each ejecta mass shell $m_i$
and mixed all mass shells within the range [$m_i$, $m_i+\delta m$].
In all but one model, we adopted $\delta m=$\,4\,\msun.
In model a3m, we used $\delta m=$\,2\,\msun.
Because of the moderate or strong mixing adopted in all models, the low metallicity
of the progenitor star is largely erased since a significant amount of
intermediate mass elements (IMEs) and iron-group elements (IGEs)
are  mixed throughout the ejecta all the way to the outer layers.
Mixing takes place primarily during the early dynamical phase of the explosion and is
largely over when the reverse shock, which travels inward into the former He-core material,
dies out. We assume that this mixed composition profile is subsequently frozen in (the only
changes are associated with the local decay of unstable isotopes).

Apart from the initial \nifs\ abundance (and the different levels of mixing between models a3 and a3m), all models (except model a4he) have the same composition in mass space (this stratification differs in velocity space if the ejecta kinetic energy is different). In the outer regions of the ejecta, the mass fraction for the main species are 0.695 for H, 0.271 for He, 0.0037 for C, 0.017 for O,  0.0027 for Si, and 0.00058 for Fe (N has a negligible mass fraction). In model a4he, the surface mass fractions for H and He are 0.457 and 0.509. With this set, we can explore the effects of ejecta kinetic energy (a2, a3, a4, a5; models also differ in \iso{56}Ni mass), \iso{56}Ni mass (for about the same ejecta mass and kinetic energy; a3 and a3ni; a4 and a4ni), \iso{56}Ni mixing (a3 and a3m), and H/He abundance ratio (a4 and a4he).

\begin{table*}
\caption{
Ejecta properties for our model set used as initial conditions for the \cmfgen\ calculations.
The last column gives the initial \nifs\ mass (prior to any radioactive decay).
\label{tab_sum}
}
\begin{center}
\begin{tabular}{
l@{\hspace{3mm}}c@{\hspace{3mm}}c@{\hspace{3mm}}
c@{\hspace{3mm}}c@{\hspace{3mm}}c@{\hspace{3mm}}
c@{\hspace{3mm}}c@{\hspace{3mm}}c@{\hspace{3mm}}
}
\hline
Model & $M_{\rm ejecta}$  & $E_{\rm kin}$ &   H & He & O & Si & Ca & \iso{56}Ni$_0$ \\
     &   [\msun]   & [10$^{51}$\,erg] &   [\msun] &   [\msun] &   [\msun] &   [\msun] &   [\msun] &   [\msun] \\
\hline
a2 &    14.30 &         0.47 &         7.35 &         5.01 &         1.00 &         0.16 &        0.017 &     0.0276 \\
a3 &     13.53 &         0.87 &         7.32 &         4.59 &         0.83 &         0.13 &        0.013 &     0.0469 \\
a3ni &  13.36 &         0.86 &         7.53 &         4.41 &         0.71 &         0.11 &        0.011 &     0.0788  \\
a3m  & 12.54 &         0.87 &         7.40 &         3.89 &         0.64 &         0.10 &        0.010 &     0.0365 \\
a4   &   13.22 &         1.26 &         7.31 &         4.40 &         0.76 &         0.12 &        0.011 &     0.0843  \\
a4he & 13.22 &         1.26 &         4.87 &         6.84 &         0.76 &         0.12 &        0.011 &     0.0788  \\
a4ni  & 12.58 &         1.25 &         7.18 &         4.02 &         0.62 &         0.10 &        0.009 &     0.2121 \\
a5     &  13.10 &         2.46 &         7.36 &         4.29 &         0.68 &         0.14 &        0.008 &     0.1571 \\
\hline
\end{tabular}
\end{center}
\end{table*}

\begin{figure}
   \includegraphics[width=\hsize]{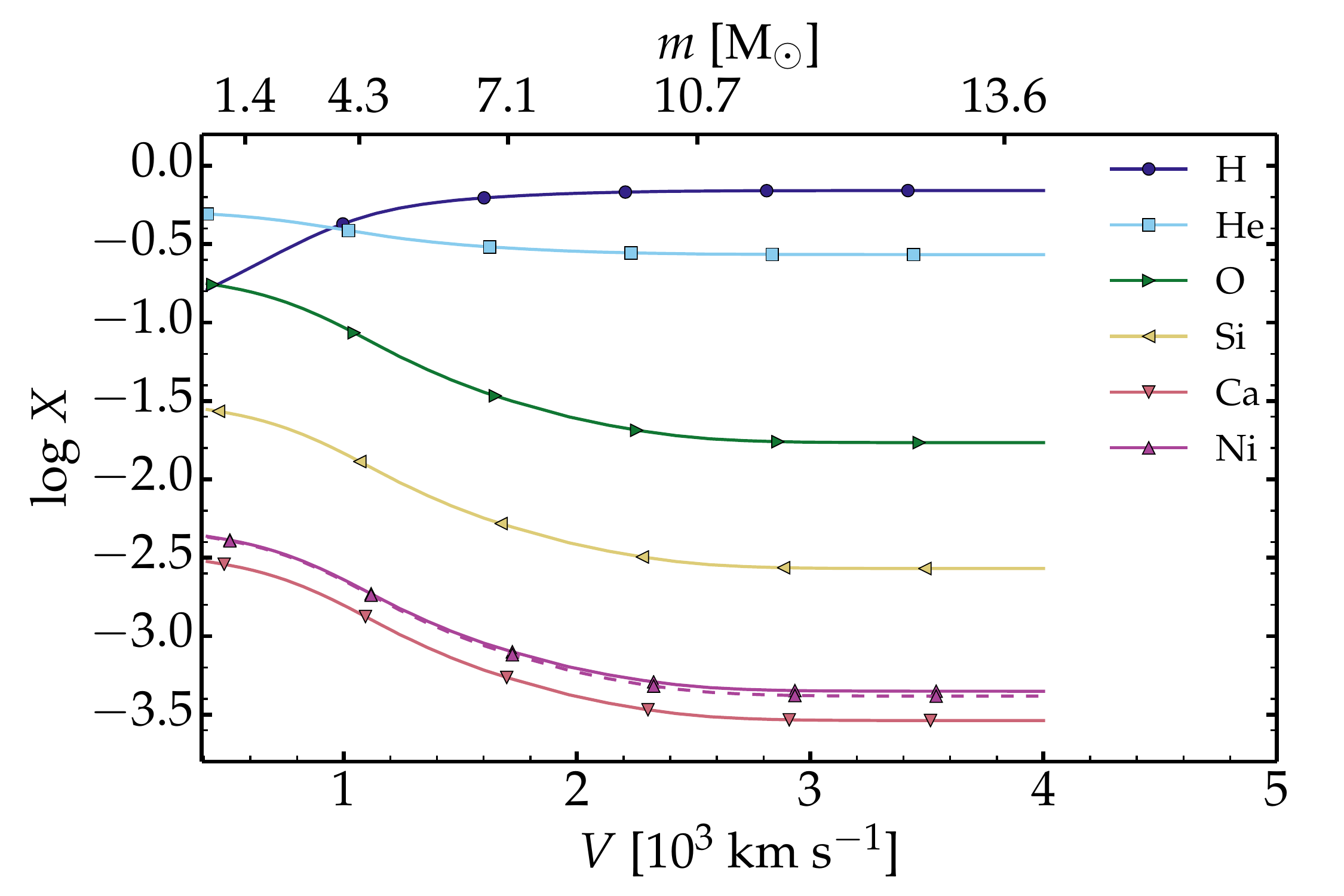}
\caption{
Composition profiles for the inner ejecta of model a2 used in the initial \cmfgen\ model
at 1.1\,d (the dashed curve corresponds to \nifs).
The outer ejecta has the same composition as the one at 4000\,\kms.
The total ejecta mass for model a2 is 14.30\,\msun.
Other models have similar profiles because they all stem from the same progenitor.
Differences arise from differences in explosion energy, \iso{56}Ni yield, and mixing.
Hence, other models show composition profiles that are shifted horizontally
(i.e., in velocity space) or vertically (i.e., in yield, which is most relevant for \iso{56}Ni,
or for H and He in model a4he).
(see Section~\ref{sect_expl} for discussion).
 \label{fig_a2_comp}
}
\end{figure}

\subsection{Radiative transfer models with \cmfgen}

At 1\,d after the explosion trigger, the ejecta are all close to homologous expansion.
The \v1d\ simulations are then remapped into \cmfgen\
\citep{HM98,DH05a,D08_time, HD12,d13_sn2p}. To avoid repetition, we refer the
reader to these papers for details about the numerical approach and setup.

In \v1d, the smoothing applied at 5\,s erases the sharp variations in composition.
However, strong density variations are present in the models at 1\,d (caused by the density
contrast between progenitor He core and H-rich envelope, which is exacerbated
by the reverse shock that it causes).
We apply a gaussian smoothing to damp these strong variations in density.
In the regions interior to
the inner boundary, we use a linear extrapolation for the density.
Because the density steeply rises at the base (the more so in the weaker explosion models),
it alters the resulting ejecta mass.
Hence, the \cmfgen\ ejecta properties differ from those obtained with \v1d.
This primarily concerns the innermost layers at the lowest velocities, and thus affects
the ejecta mass (by up to 7\,\%), the scaled \nifs\ mass (by up to 25\%),
and the kinetic energy aimed for (by up to $\sim$\,15\,\%, which arises
mostly because of the change in ejecta mass).
Since the offset in ejecta mass stems mostly from a change in density in the
innermost ejecta layers rich in metals, we suspect the impact on the light curve will
be modest.
The ejecta properties used for our simulations with \cmfgen\ are presented
in Table~\ref{tab_sum} and the ejecta composition profile for model a2 is
shown in Fig.~\ref{fig_a2_comp}.

The low nitrogen content in the progenitor  and in our ejecta (about
0.0001\,\msun) is a result of using a very low metallicity progenitor. But nitrogen
has no influence on the SN light curves and no nitrogen line is observed in SNe II-pec.
Such limitations have little impact on our results. Instead, with strong mixing, the
spectrum forms in a metal-rich environment a few weeks after explosion, irrespective
of the adopted progenitor metallicity.

  Metals like Ba, not included in \mesa\ nor in \v1d, are given a constant mass fraction equal
  to the corresponding LMC metallicity value.
  Metals between Ne and Ni that are included in \mesa\ and \v1d are given a minimum
  mass fraction equal to the corresponding LMC metallicity value (this applies to the most
  abundant stable isotope for each element).
  The LMC value is adopted here since it is the metallicity inferred for the environments
  of Type II-pec SNe \citep{taddia_2pec_z_13}.

    The numerical setup is comparable to that of \citet{d13_sn2p}. We use the same
    model atoms (with the addition of Ba), with updates to the atomic data (in particular
    for Fe and Co) as described in \citet{d14_tech}.
    We treat the following ions:  H\one, He\one-\two, C\one-\three, N\one-\two,
    O\one-\three, Ne\one-\two, Na\one, Mg\one-\three, Si\one-\three, S\one-\three,
    Ar\one-\two, K\one, Ca\one-\three, Sc\one-\three, Ti\two-\three, Cr\two-\four,
    Fe\one-\five, Co\two-\five, Ni\two-\five, and Ba\one--\two.

   In our \cmfgen\ simulations, the intrinsic line width is set by the effective thermal
   velocity $V_{\rm th}$,
   which is controlled by the local gas temperature
   $T$, the atomic mass $A$ of the element or ion, as well as the adopted micro-turbulent velocity
   of the medium $V_{\rm turb}$. We use the definition
   \begin{equation}
   V_{\rm th}^2 =  k_{\rm B} T /Am_{\rm u} +   V_{\rm turb}^2    \, ,
   \end{equation}
   where $m_{\rm u}$ is the atomic mass unit and $k_{\rm B}$ is the Boltzmann
   constant.
   If the turbulent velocity is zero, the line width at 10,000\,K
   decreases from about 10\,\kms\ for H\one\ lines down to about 1\,\kms\ for
   an Fe line.
   The broader is the intrinsic width of lines, the smaller is the total number of frequency
   points used in \cmfgen\ (we use a fixed number of frequency points, typically
   $5-6$, to resolve a given line; line overlap does not increase the number of allocated frequencies).
In the present set of simulations, the number of frequency points is $\sim$\,100,000 for
$V_{\rm turb}=$\,50\,\kms\ and $\sim$\,500,000 for $V_{\rm turb}=$\,10\,\kms.
The computational cost of running \cmfgen\ with a small value of $V_{\rm turb}$
is non trivial.

\begin{figure*}
   \includegraphics[width=0.5\hsize]{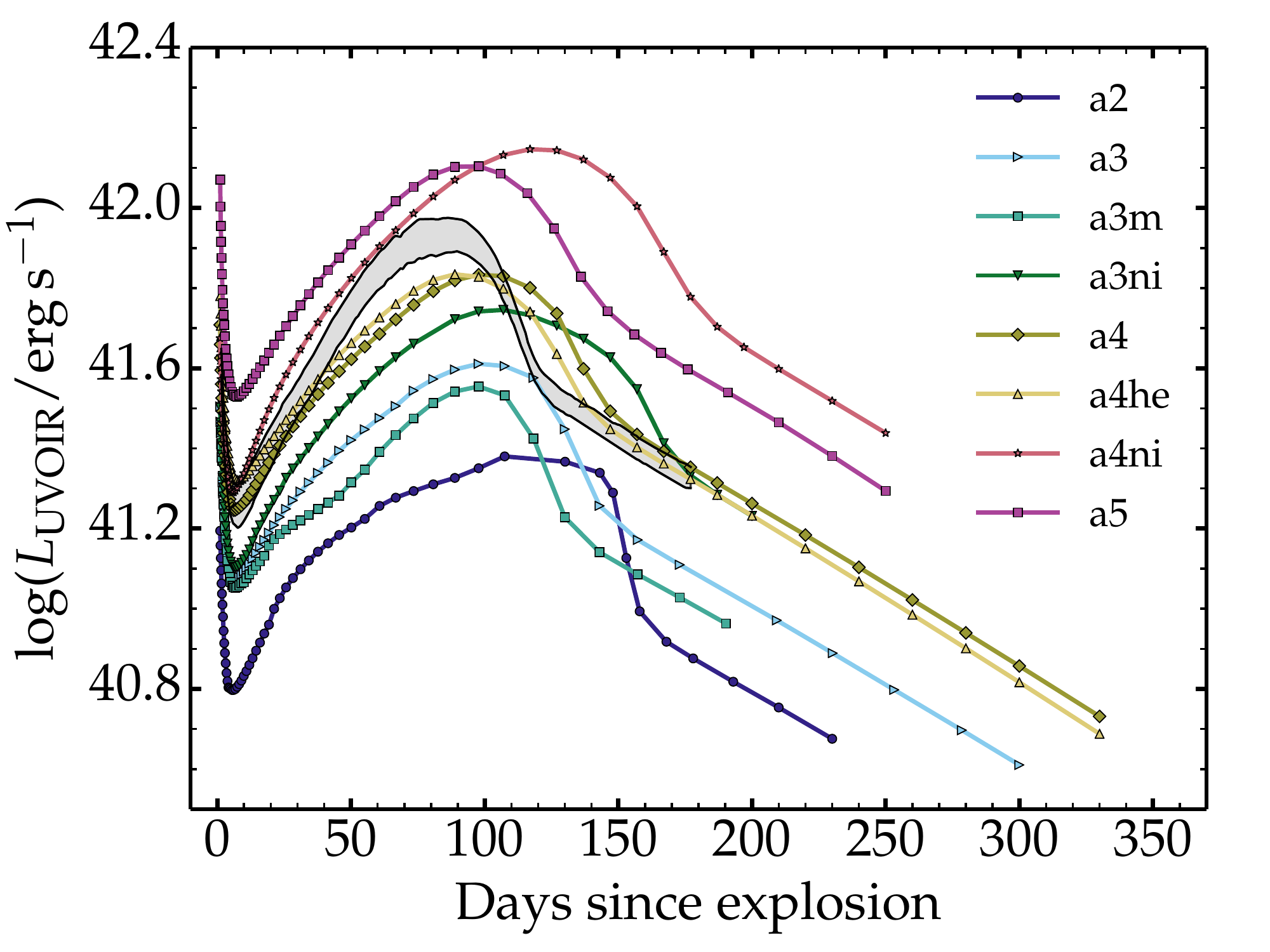}
   \includegraphics[width=0.5\hsize]{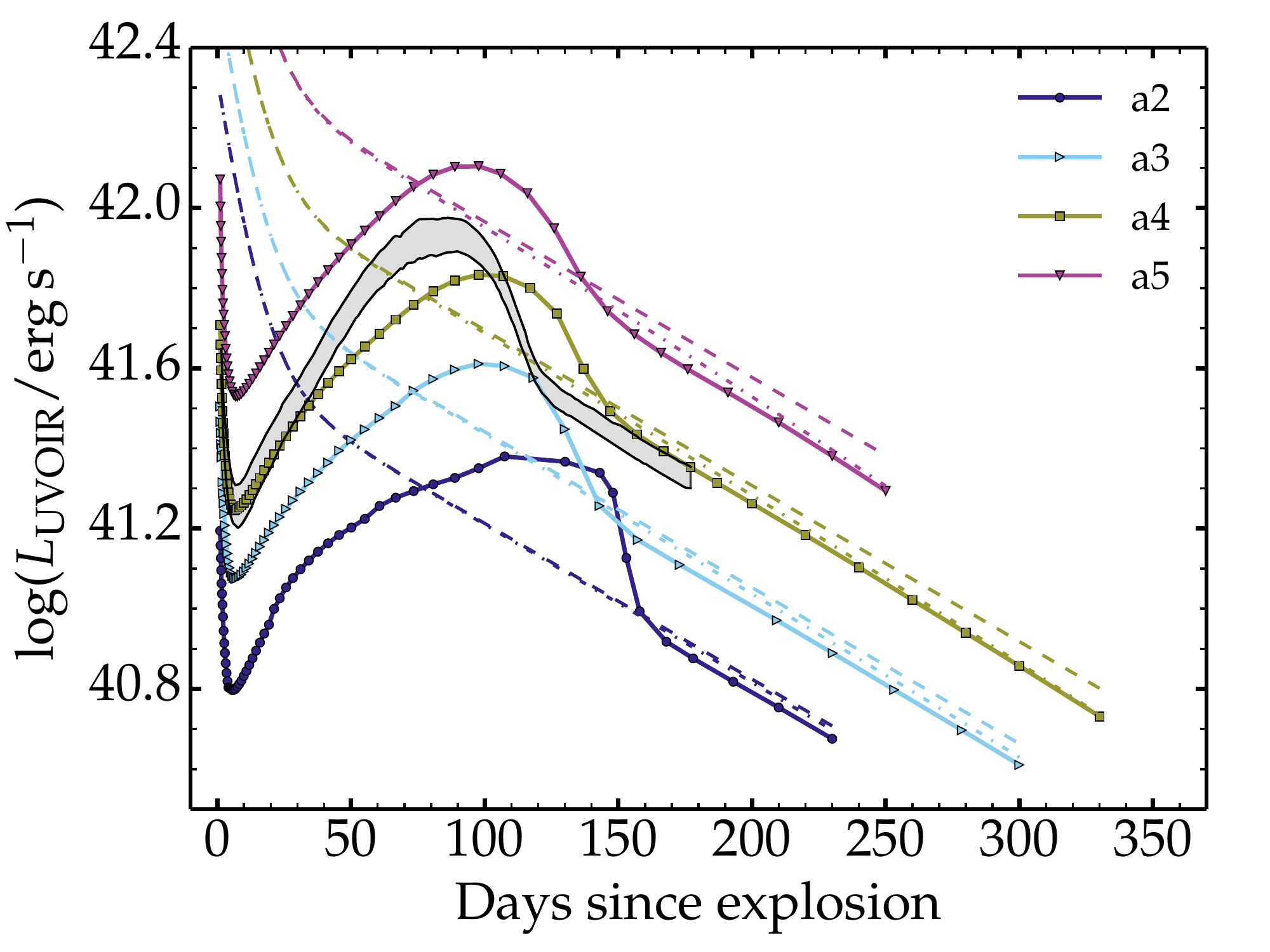}
\caption{
Left: UVOIR  light curves
for our \cmfgen\ simulations, together with
the inferred light curve for SN\,1987A (the shaded area is bounded by the
UVOIR light curves inferred by \citealt{catchpole_87A_87} and \citealt{hamuy_87A_88}).
Right: Same as left, but now for the simulations a2, a3, a4, and a5 (which differ primarily
in ejecta kinetic energy and \nifs\ mass).
We also show the total decay power emitted (dashed line) and absorbed (dash-dotted line).
and in \nifs\ mass).
 \label{fig_lbol}
}
\end{figure*}

\begin{figure*}
   \includegraphics[width=0.5\hsize]{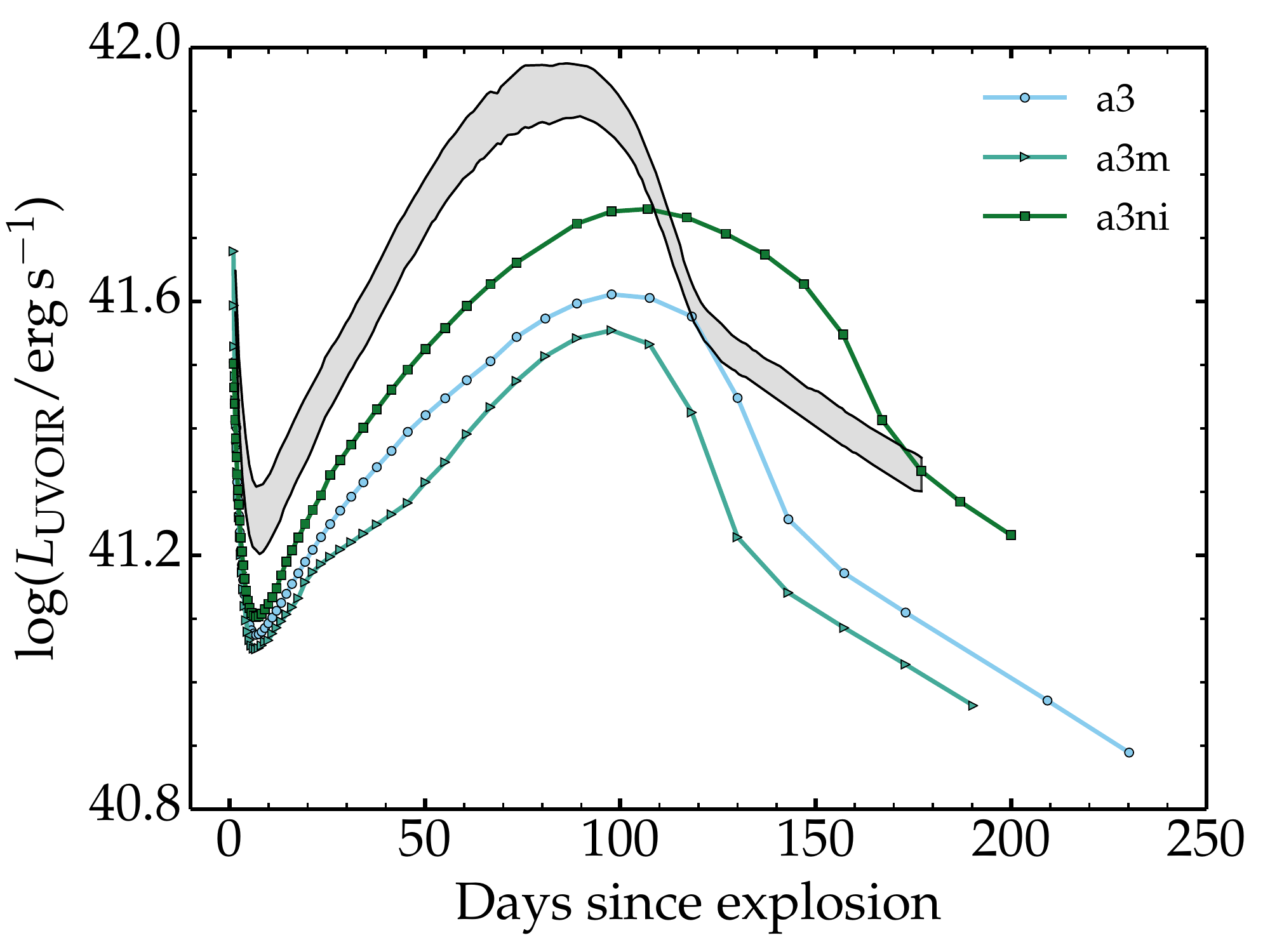}
   \includegraphics[width=0.5\hsize]{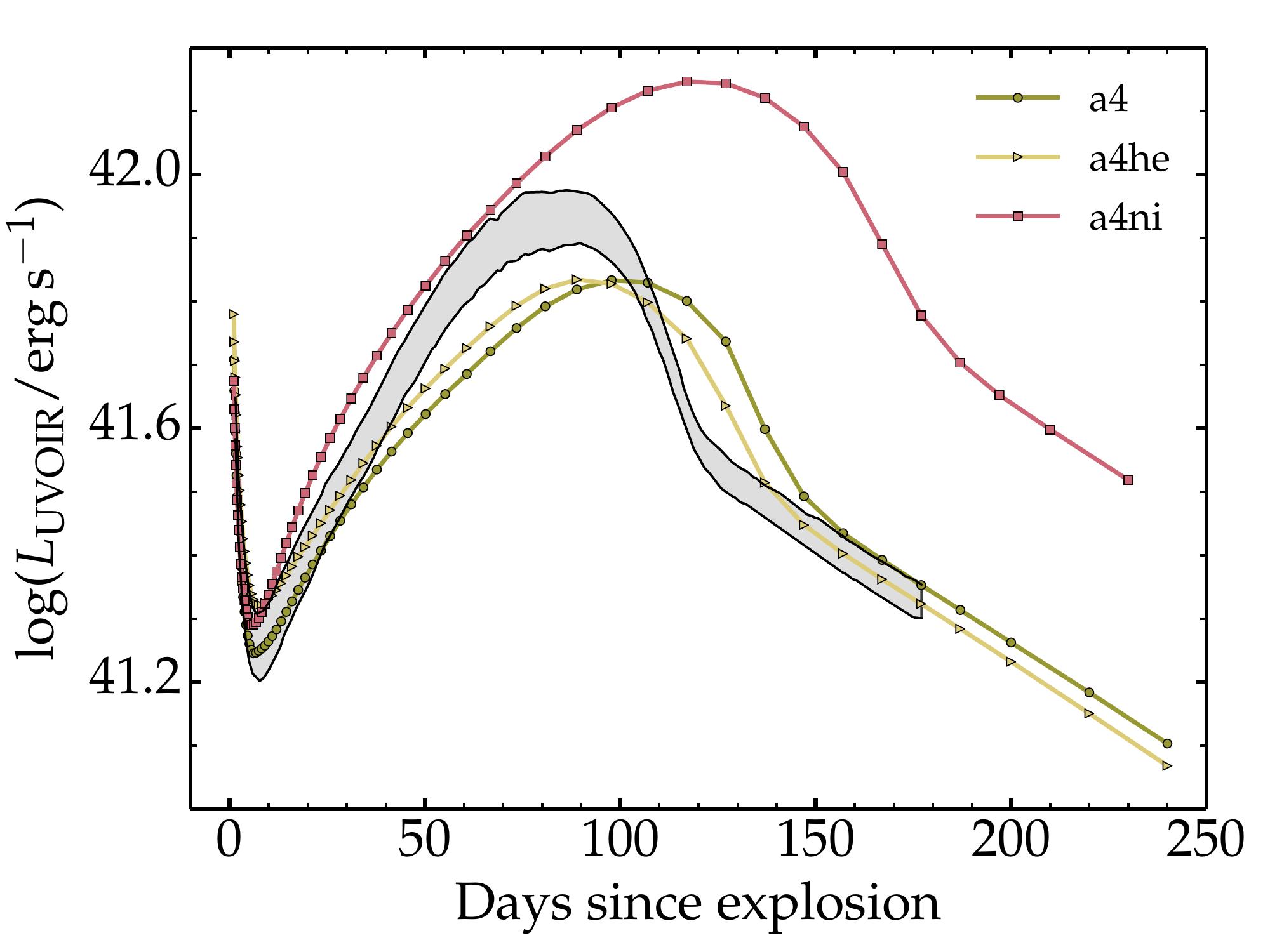}
\caption{
Left: Same as Fig.~\ref{fig_lbol}, but now for the simulations
a3, a3m, and a3ni (which differ in chemical mixing magnitude and in \nifs\ mass).
Right: Same as left, but now for the simulations
a4, a4he, and a4ni (which differ in H/He ratio in the progenitor H-rich envelope
and in \nifs\ mass).
 \label{fig_lbol_bis}
}
\end{figure*}

   Exceptionally, for this study, all simulations were performed with a turbulent
   velocity of 10\,\kms, except for the low energy model a2. The simulations for this
   ejecta model were taking an eternity to converge with 10\,\kms\
   and so were stopped after 37 time steps, at a SN age of 31.14\,d -- the time sequence
   for model a2 is thus performed with a turbulent velocity of 50\,\kms.
   Model a4 was run both with 10\,\kms\ and 50\,\kms\ (the latter corresponds
   to model Bsm in \citealt{d18_fcl}). Hence, for models a2 and a4, we have two sequences
   (truncated at 31.14\,d for a2) performed with a turbulent velocity of 10\,\kms\
   and 50\,\kms. Comparison of the two reveal that the bolometric light curves
   are identical to within 1\,\% at all times and the optical colors differ modestly.
   Simulations performed with the lower value of $V_{\rm turb}$ have bluer optical colors,
   typically by $\lesssim$\,0.1\,mag. This is caused by the enhanced line blanketing
   induced by a larger value of $V_{\rm turb}$. The models with a greater
   turbulent velocity have a larger Rosseland mean optical depth but this
   does not impact the light curve. The reason is that there is effectively no diffusion
   from large optical depth. What matters is diffusion through the photospheric layers,
   and for that part, the turbulent velocity seems to impact the redistribution of flux
   (the color) rather than its escape (the total flux). So, the use of a larger value
   of $V_{\rm turb}$ has little impact on the current results.

    The present simulations include only the radioactive decay
    of \iso{56}Ni and \iso{56}Co. All other isotopes are considered stable (most are).
    We compute the non-local energy deposition at all epochs using a purely
    absorptive treatment and a gray opacity of $0.06 Y_{\rm e}$\,cm$^2$\,g$^{-1}$
    ($Y_{\rm e}$ is the electron fraction). We treat the associated non-thermal processes
    within the non-LTE time-dependent solver of \cmfgen\
    (for details, see \citealt{d12_snibc} and \citealt{li_etal_12_nonte}).

While preparing this article, \citet{d18_fcl} published a study on the effect of clumping
in Type II SNe during the photospheric phase. Using a volume filling factor approach,
and ignoring any impact from porosity and chemical segregation, they find that clumping
enhances the recombination rate, causing the photosphere to recede faster through the
ejecta and boosting the luminosity. This clumping effect leads to a shorter rise time to
maximum and an earlier transition to the nebular phase. In other words, all else being
the same, a clumped ejecta  tends to behave as a smooth ejecta of a lower mass.
Clumping is ignored in this study.

With \cmfgen, doing the eight time sequences presented here was a major endeavor. Each time sequence represents 65 time steps, each taking between two days and a week of computing. The total number of models or epochs is 520 and required more than six months of computing. Doing more with our current super-computer resources is not possible.

\begin{figure}
\includegraphics[width=\hsize]{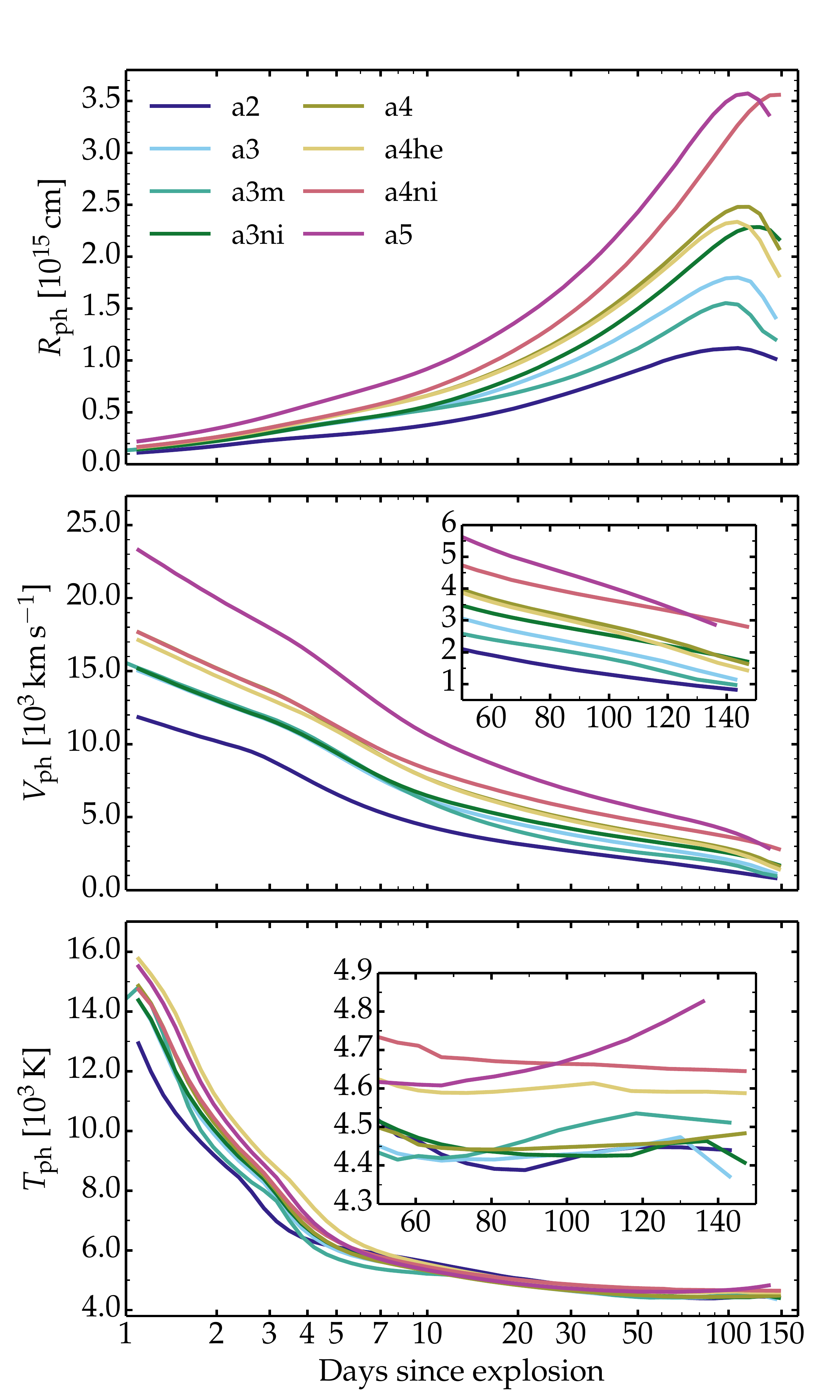}
\caption{Evolution of the radius, velocity (which can also be deduced directly from
the radius and the time elapsed since explosion), and temperature at the photosphere
for our set models. The insets in the lower two panels zoom on the epoch around
bolometric maximum, when the differences in velocity and temperature between models
are harder to discern.
 \label{fig_phot_prop}
}
\end{figure}

\begin{figure}
   \includegraphics[width=\hsize]{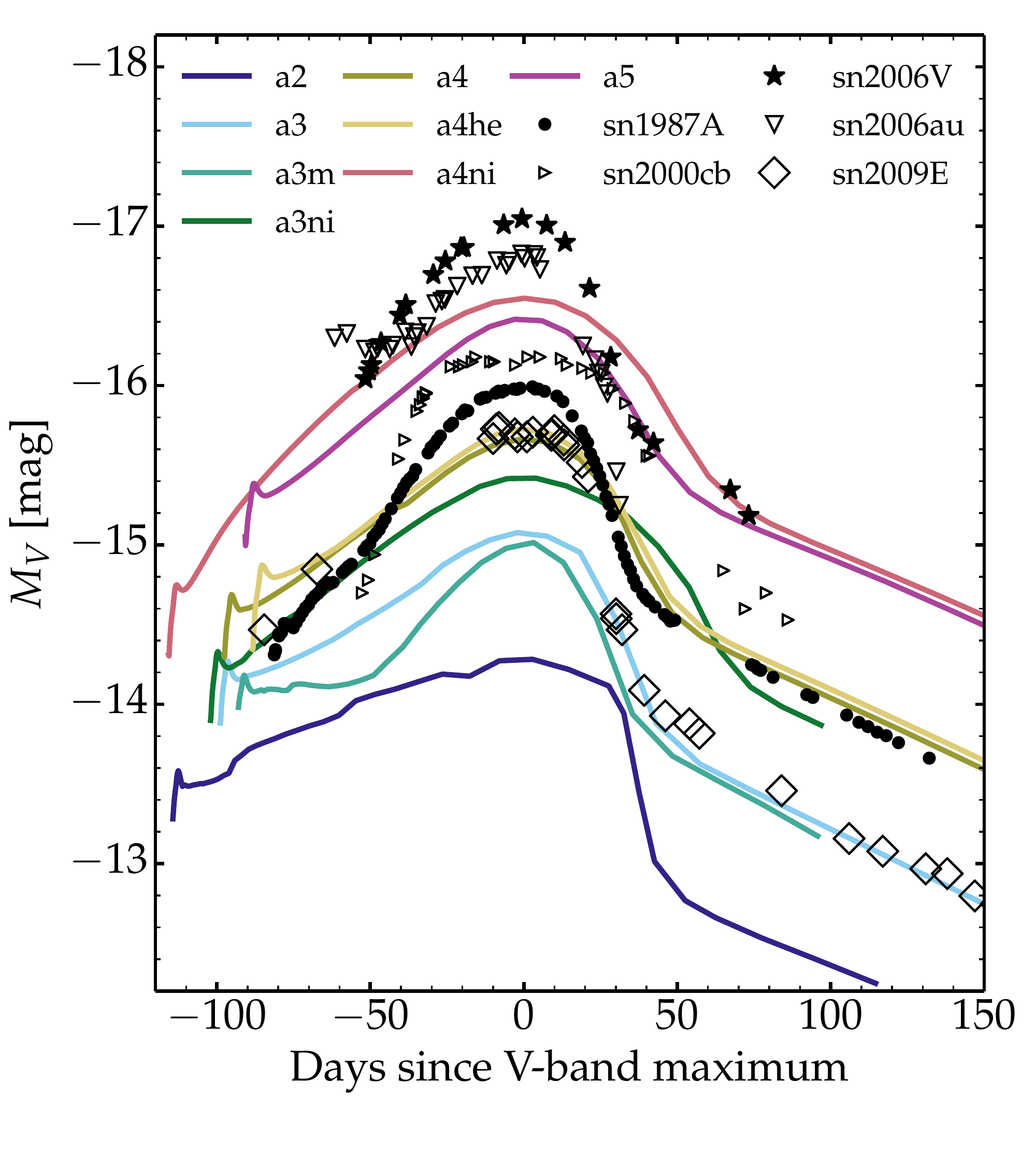}
\caption{Comparison of the $V$-band light curves for observed SNe II-pec
and for our set of models (the time axis corresponds to the time elapsed
since $V$-band maximum; this time is not accurately known given the
width of the light curves and the noise in the observed photometry).
If we were to use a negligible reddening for SN\,2006au (see Sections~\ref{sect_obs_06au}
and \ref{sect_spec_06au}), its optical brightness and color would be comparable to those of SN\,1987A.
The $V$-band light curve of SN\,2009mw is very similar to that of SN\,2009E
and thus not shown \citep{takats_09mw_16}.
The SN properties used to make this figure are listed in Table~\ref{tab_obs}.
\label{fig_nlc_V}
}
\end{figure}

\section{UVOIR Light curves}
\label{sect_luvoir}

Figures~\ref{fig_lbol} and \ref{fig_lbol_bis}
show the UVOIR light curve for our set of models.\footnote{We define the UVOIR flux
as the flux falling between 1000\,\AA\ and 2.5$\mu$m.}
 In the left panel of Fig.~\ref{fig_lbol},
we show the whole set of models together with the inferred UVOIR
light curve of  SN\,1987A \citep{catchpole_87A_87,hamuy_87A_88}.
All models  exhibit a similar light curve morphology, with a $\sim$\,100\,d rise
to a $\sim$\,50\,d broad maximum. In all cases the rise time is longer than
observed for SN\,1987A, and the width of the maximum is also broader.
The brightening rate on the way to maximum is lower than for SN\,1987A
unless the \nifs\ mass is well in excess of the value inferred for SN\,1987A
(which is about 0.08\,\msun, depending on the adopted reddening and
the amount of flux not accounted for).

In the right panel of Fig.~\ref{fig_lbol}, we only show the set of models a2, a3, a4,
and a5, which differ primarily in ejecta kinetic energy and \nifs\ mass.
However, we now add the total decay power emitted and absorbed.
The difference between the latter two arises from $\gamma$-ray escape.
The weaker explosions trap more efficiently $\gamma$-rays. For the more
energetic explosions corresponding to models a4 and a5, $\gamma$-ray leakage
starts at or soon after maximum.
The difference between the total decay power absorbed and the UVOIR luminosities
at nebular times arises from the flux not accounted for that falls beyond 2.5\,$\mu$m
(at earlier times the luminosity is affected by optical-depth effects).
As time progresses in the nebular phase, the UVOIR luminosity converges toward the total decay power absorbed (or the bolometric luminosity).

We find that the bolometric luminosity at maximum is typically $40-60$\,\% larger
than the total decay power absorbed at that time. The SN\,1987A light curve
shows a similar offset.

The slopes at nebular times all follow closely the \cofs\ decay slope for full trapping.
In models a2$-$a3, this arises because full trapping holds.
In models a4$-$a5, it arises because the increasing $\gamma$-ray escape fraction is
compensated by the enhanced fraction of the flux falling between 1000\,\AA\ and
2.5\,$\mu$m (that fraction recorded by the UVOIR luminosity).

\begin{figure}
   \includegraphics[width=\hsize]{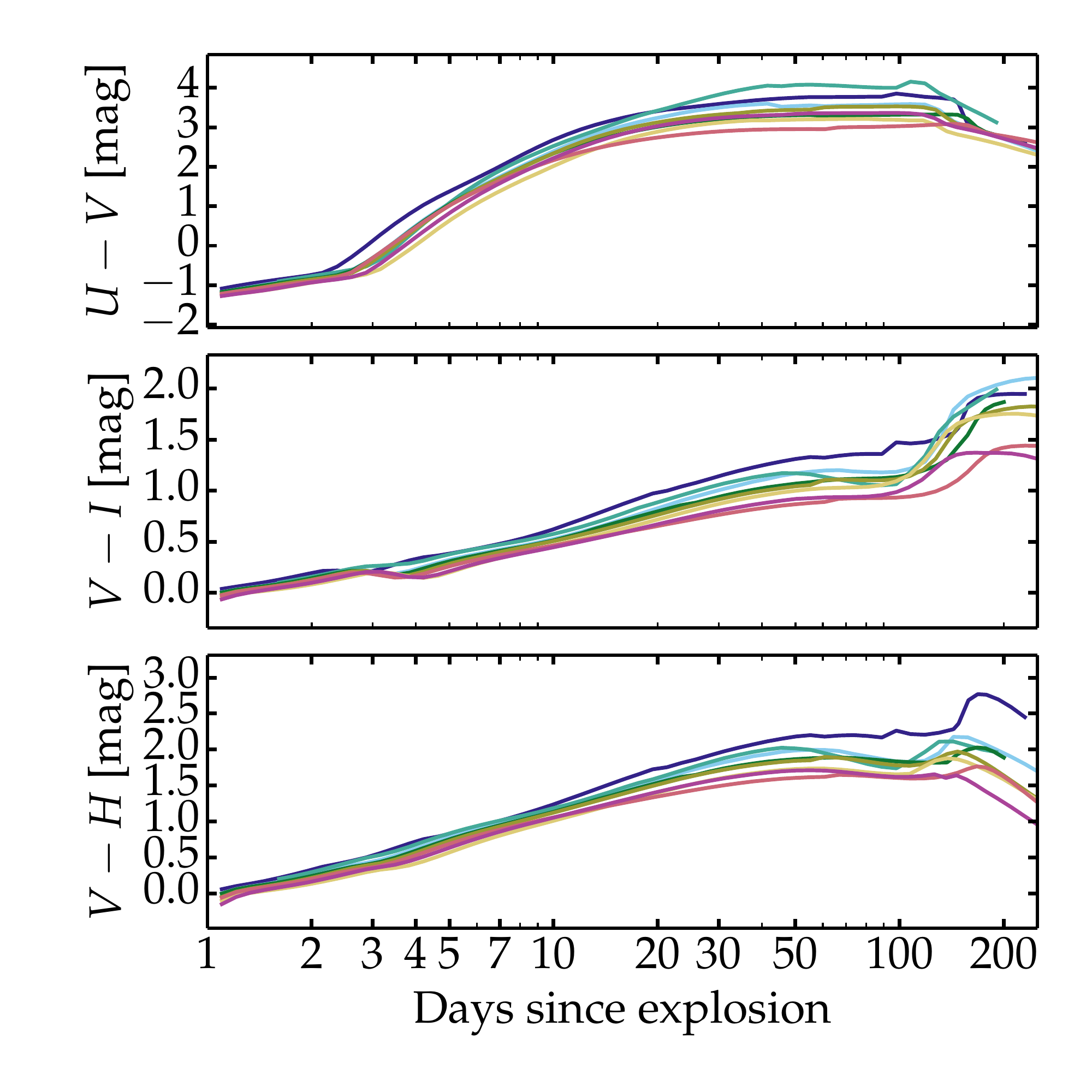}
\caption{
Evolution of the $U-V$, $V-I$, and $V-H$ colors for our set of models.
\label{fig_color}
}
\end{figure}

\begin{figure}
   \includegraphics[width=\hsize]{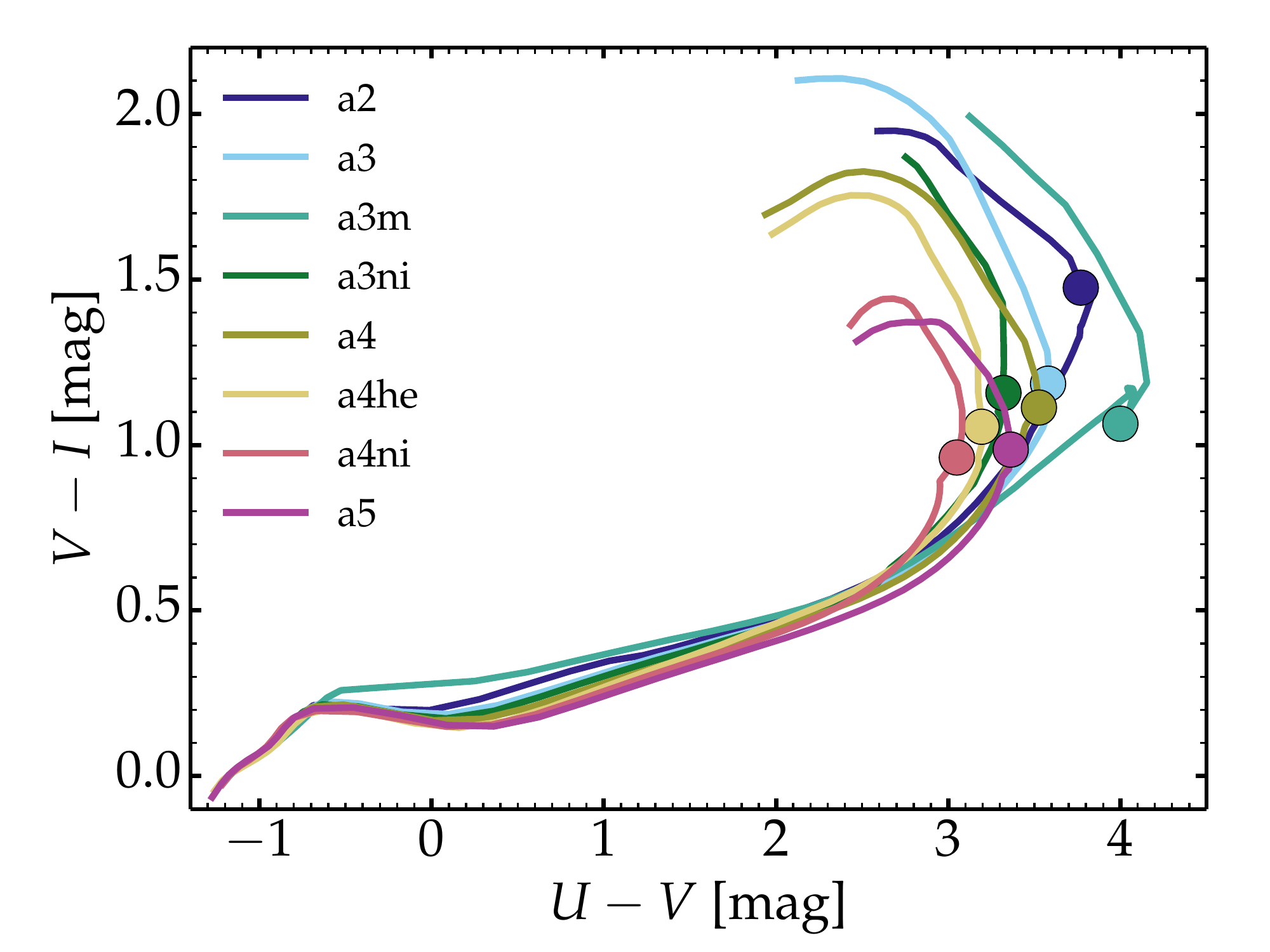}
\caption{
Color-color diagram for our set of models. The filled dot indicates
the time of bolometric maximum -- all models start in the bottom-left
corner with a blue optical color.
\label{fig_col_col}
}
\end{figure}

These variations in brightness are in part reflected by the evolution of the
conditions at the photosphere (Fig.~\ref{fig_phot_prop}).
The initial drop in luminosity stops at about 5\,d when the photospheric
temperature stabilizes at the H recombination temperature.
Then, as the photospheric radius increases, the luminosity also increases,
reaching a maximum when the photospheric radius is also maximum
(having values in the range $1-3.5 \times 10^{15}$\,cm).

\begin{figure*}
   \includegraphics[width=0.33\hsize]{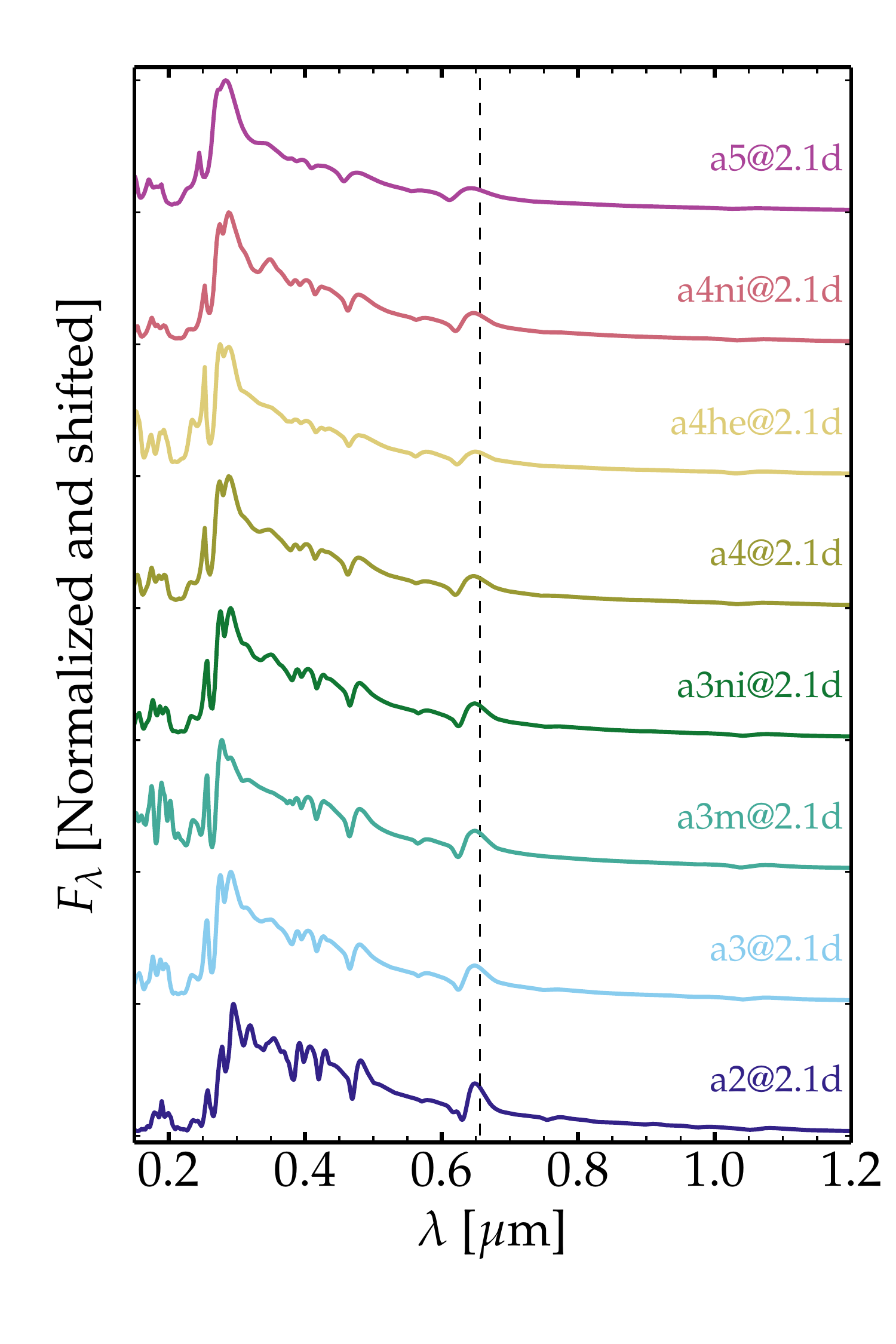}
   \includegraphics[width=0.33\hsize]{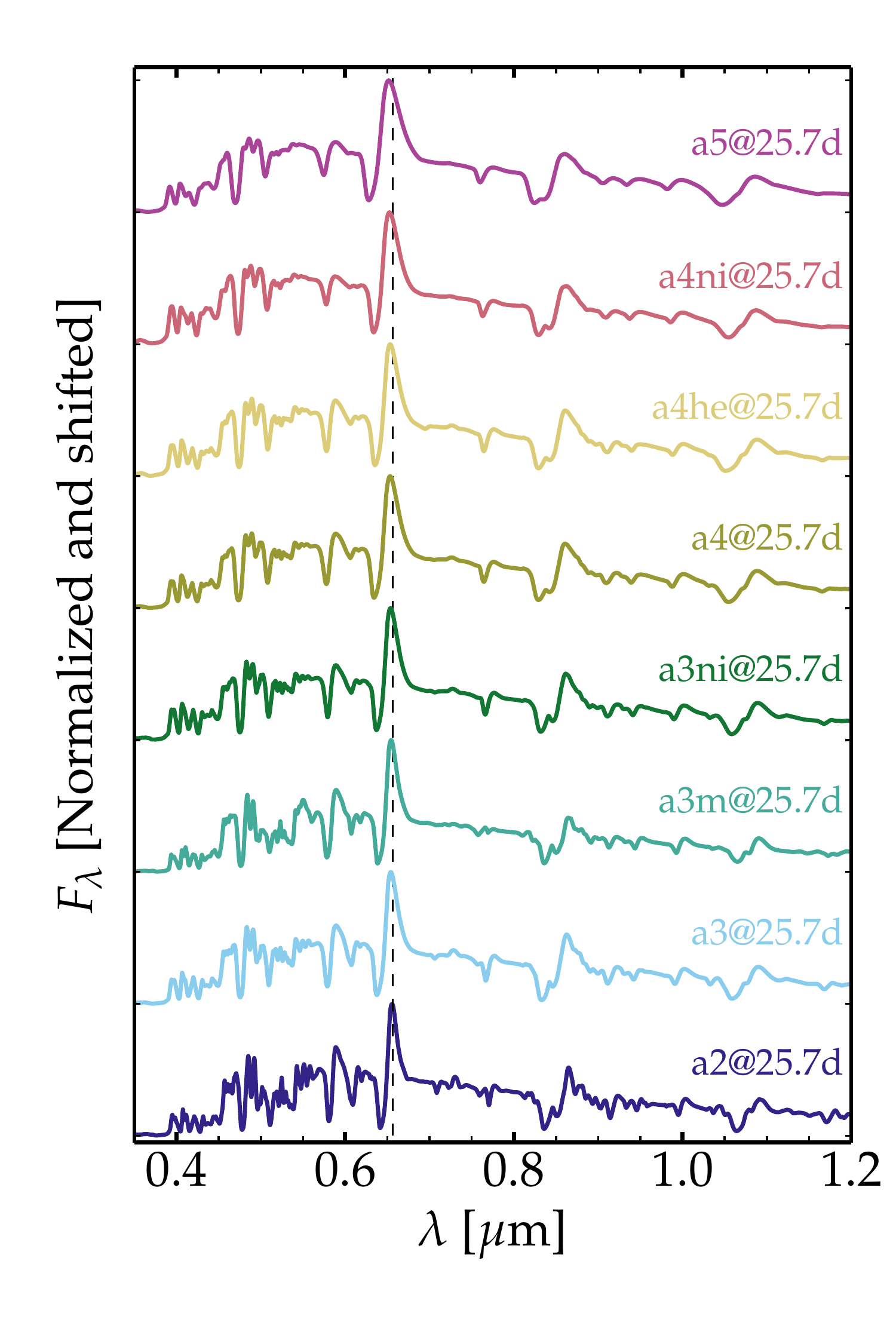}
   \includegraphics[width=0.33\hsize]{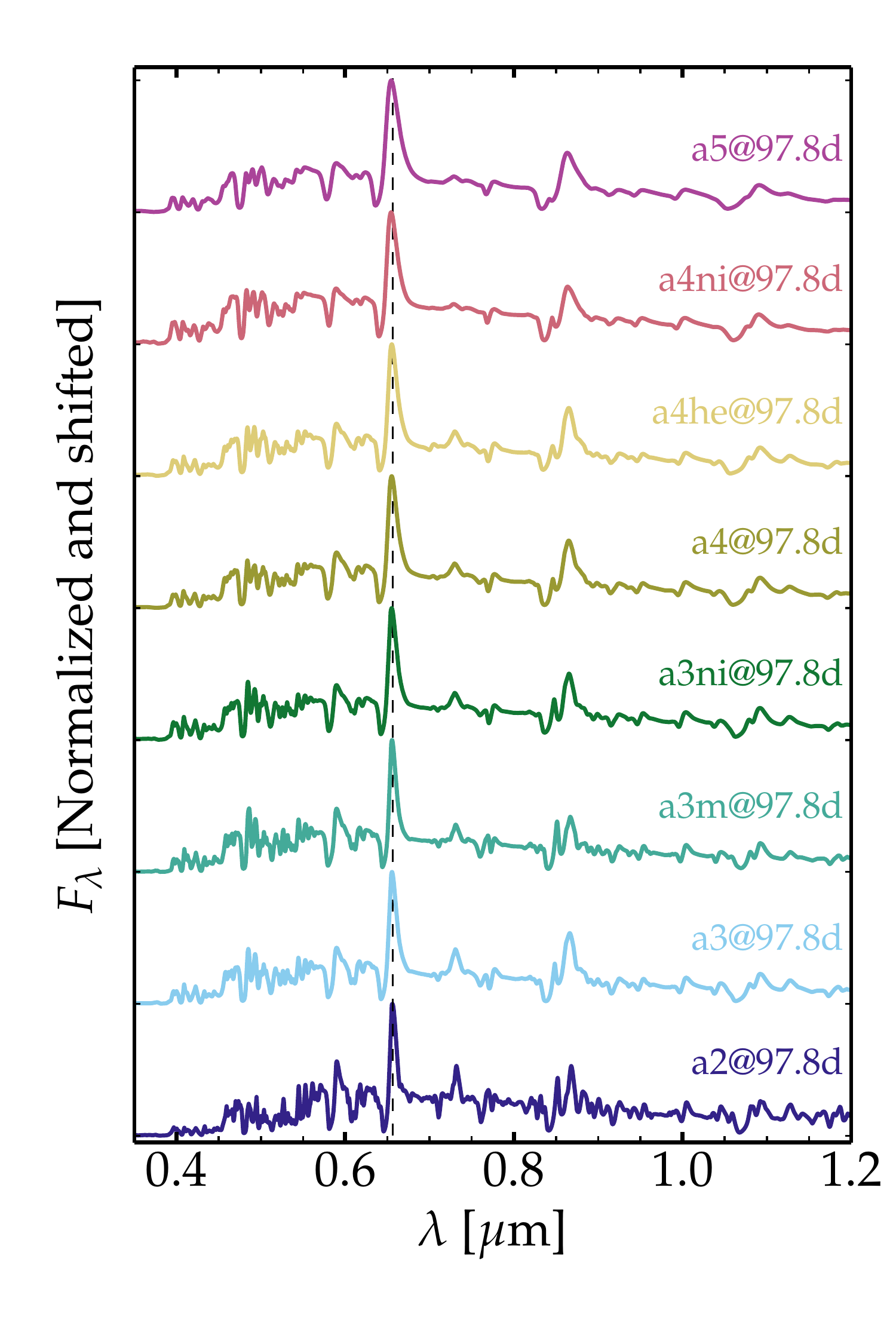}
\caption{
Spectral montage for our set of models at 2.1\,d (left),  25.7\,d (center),
and 97.8\,d (right) after explosion (i.e., around the time of bolometric maximum).
Each spectrum has been normalized and shifted (a tickmark gives the
zero-flux level for each spectrum).
The vertical dashed line locates the rest wavelength of H$\alpha$
and reveals the blue shift of the emission peak at earlier times
and higher explosion energies.
\label{fig_spec_3epochs}
}
\end{figure*}

Figure~\ref{fig_lbol_bis} shows two subsets of models having the same ejecta mass
and kinetic energies but differing in composition. Models with enhanced \nifs\ are
more luminous at all times than their counterpart with less \nifs.
At the earliest times, the offset is small because \nifs\ decay power has only a small influence
on the outer layers from where the radiation escapes. The offset is however not zero because
the mixing is strong in our ejecta models. For example, model a3m is fainter than model
a3 because it has less \nifs\ and weaker \nifs\ mixing (the ejecta mass is also lower by 1\,\msun).
Model a3ni is more luminous than models a3 and a3m
because it has twice as much \nifs, which is also strongly mixed.
These dependences are well understood and have been discussed extensively in
the past (see, e.g, \citealt{shigeyama_nomoto_90} or \citealt{blinnikov_87A_00}).
Model a3 and its variants do not match well the UVOIR light curve of SN\,1987A,
but the spectral evolution of model a3 is a good match to SN\,1987A beyond $\sim$\,15\,d.

In the right panel of Fig.~\ref{fig_lbol_bis}, we show the UVOIR light curves
for model a4 and a4he. The two models have very similar \nifs\ mass (and the same ejecta mass)
but model a4he has more He and less H -- the abundance profile for He and H are
qualitatively the same in both models, except that the profile for He (H) is shifted to
a greater (lower) mass fraction in model a4he (see for reference the composition shown
in Figs.~\ref{fig_prog} and \ref{fig_a2_comp}). The composition profile of model a4he is more
characteristic of the RSG composition of the models produced in \citet{d13_sn2p}. The comparison
of models a4 and a4he shows that the He content of the envelope matters --
see, e.g., Fig.~6 of \citet{KW09} for a similar study in the context of Type II-P SNe.
The greater the He content (or equivalently the lower the H content), the faster the rise to maximum.
The model is brighter earlier and fainter later,  essentially at constant total time-integrated
bolometric luminosity (the comparison is not exact here since the \nifs\ mass
differs by 7\% between the two models;
with the same \nifs\ mass, model a4he would have been a little brighter
earlier on, and its maximum luminosity would have been greater, increasing the
contrast with model a4). The origin of the light curve differences is that the lower
the H content, the greater the He content, the harder it is to keep the gas ionized.
Consequently, the gas at and above the photosphere recombines more efficiently,
allowing the photosphere to recede faster.
Compare to model a4, model a4he has a photosphere that is more compact
and hotter by $5-10$\,\% on the way to bolometric maximum (Fig.~\ref{fig_phot_prop}).
This speeds up the recession of the photosphere and the release of stored energy.
For the same amount of stored energy (arising from shock-deposited energy
and \nifs\ decay heating), the greater helium content shifts the light curve left ward,
acting as if the ejecta mass is smaller.
These effects are similar to those arising from clumping \citep{d18_fcl}.
With clumping, model a4he may become close to matching the SN\,1987A light curve.

Our \cmfgen\ light curves are qualitatively similar to those published before
for BSG explosions (see, e.g., \citealt{woosley_87a_88}; \citealt{shigeyama_nomoto_90};
\citealt{blinnikov_87A_00}). For example, the model 14E1 (with strong mixing) of
\citet[see their Table~2 and Fig.~16]{shigeyama_nomoto_90} appears similar to our
model a4 (or a4he) and yields a similar bolometric light curve, brighter early on, rising
slightly faster and peaking at a brighter bolometric maximum. Our models a4 and a4he
match the SN\,1987A light curve better at early times but less well around the time
of maximum (model a4he peaks about 10\,d later than SN\,1987A). As argued
in \citet{d18_fcl}, a faster rise and an earlier peak can be obtained by invoking
clumping. \citet{blinnikov_87A_00} obtain similar results to \citet{shigeyama_nomoto_90}
for the same progenitor. Numerous other studies have been published, pointing to similar
conclusions for SN\,1987A and its progenitor.

\section{$V$-band light curves and color evolution}
\label{sect_photometric}

Figure~\ref{fig_nlc_V} show the $V$-band light curve for all models.
At early times, when the photosphere of these SN ejecta models is hot, a large
fraction of the flux is emitted short ward of the optical. Hence, the  UVOIR or
the bolometric luminosity is large while the $V$-band absolute magnitude is low (faint).
As soon as the photospheric temperature stabilizes at about 4500\,K
and the SN enters the recombination phase (Fig.~\ref{fig_phot_prop}),
the UVOIR and the bolometric luminosity follow the same evolution
and peak at the same time.

Figure~\ref{fig_nlc_V} also shows the inferred absolute $V$-band light curves
for our selected sample of
Type II-pec SNe  (see also Section~\ref{sect_obs} and Table~\ref{tab_obs}).
The distribution of model and observed light curves overlap (i.e., offsets
are typically $\lesssim$\,1\,mag) , although the
models tend to be brighter early on and peak at a fainter magnitude  (the brightening
rates differ).
Our model set also extends to peak magnitudes that are
1.5\,mag fainter than observed (because we design some BSG explosions models
to have a low \nifs\ mass).
A smaller progenitor radius would reduce the luminosity early on
but it would accelerate the cooling and the brightening in the optical.
Since we have only one progenitor model, we have not tested this
with \cmfgen\ here and cannot comment on it further.
A weaker \nifs\ mixing reduces the brightening at early times
and strengthen it later, as obtained for model a3m. There is however
no observed SN in that range of brightness at maximum. Apart from SN\,1987A,
the dataset of observed SNe II-pec has usually a poor coverage
prior to $1-2$ months before maximum.

Our Type II-pec SN models with a standard explosion energy (e.g., a4) have a $V$-band brightness
of $-14$\,mag at 10\,d (not influenced by \nifs\ decay at that time), which is comparable
to low-energy explosions in RSG stars leading to a Type II-P like SN\,2008bk \citep{lisakov_08bk_17}.
In other words, a faint early optical brightness is not exclusively a sign of a small progenitor radius.

 \begin{figure}[h!]
   \includegraphics[width=0.97\hsize]{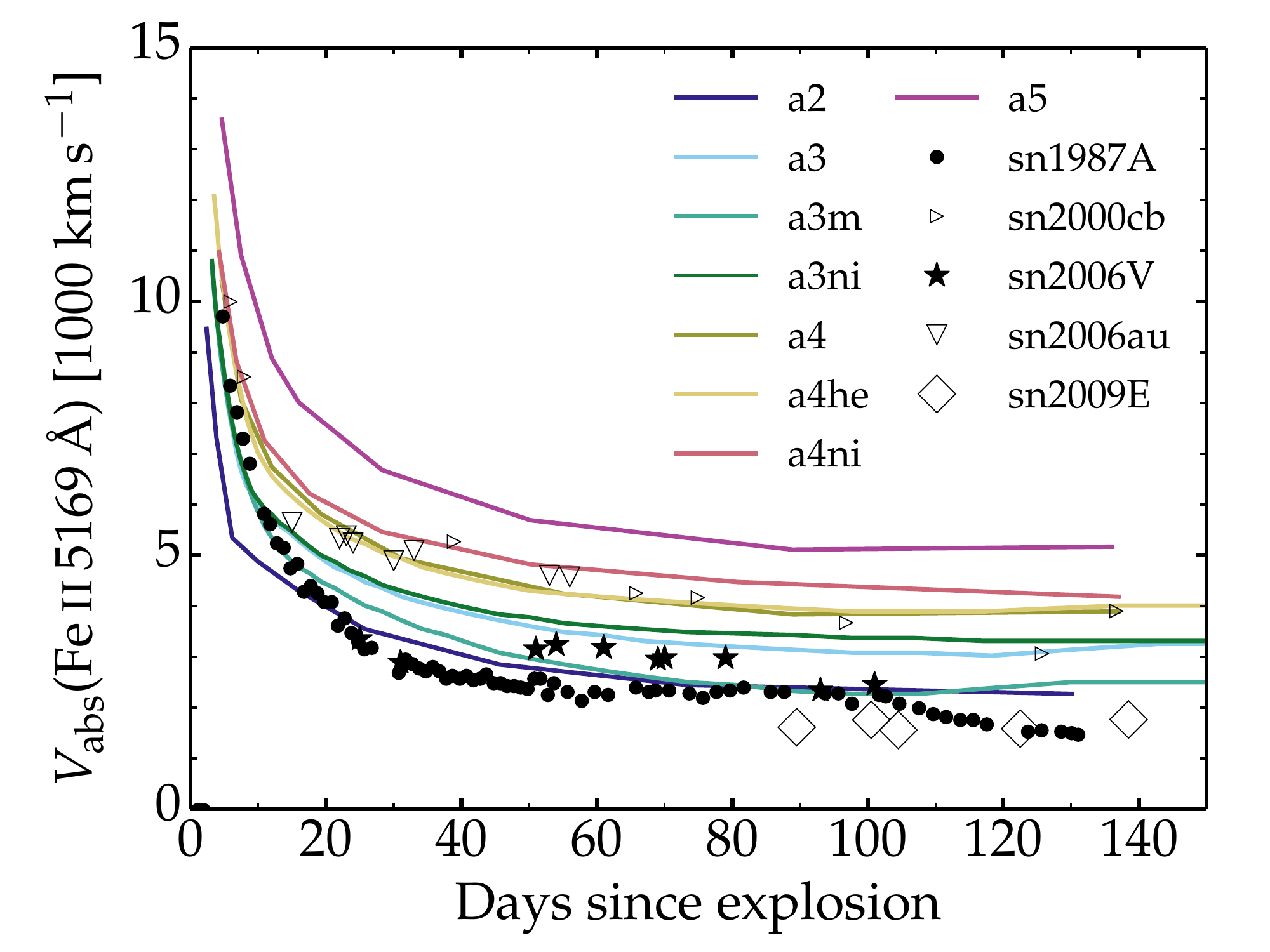}
   \includegraphics[width=0.97\hsize]{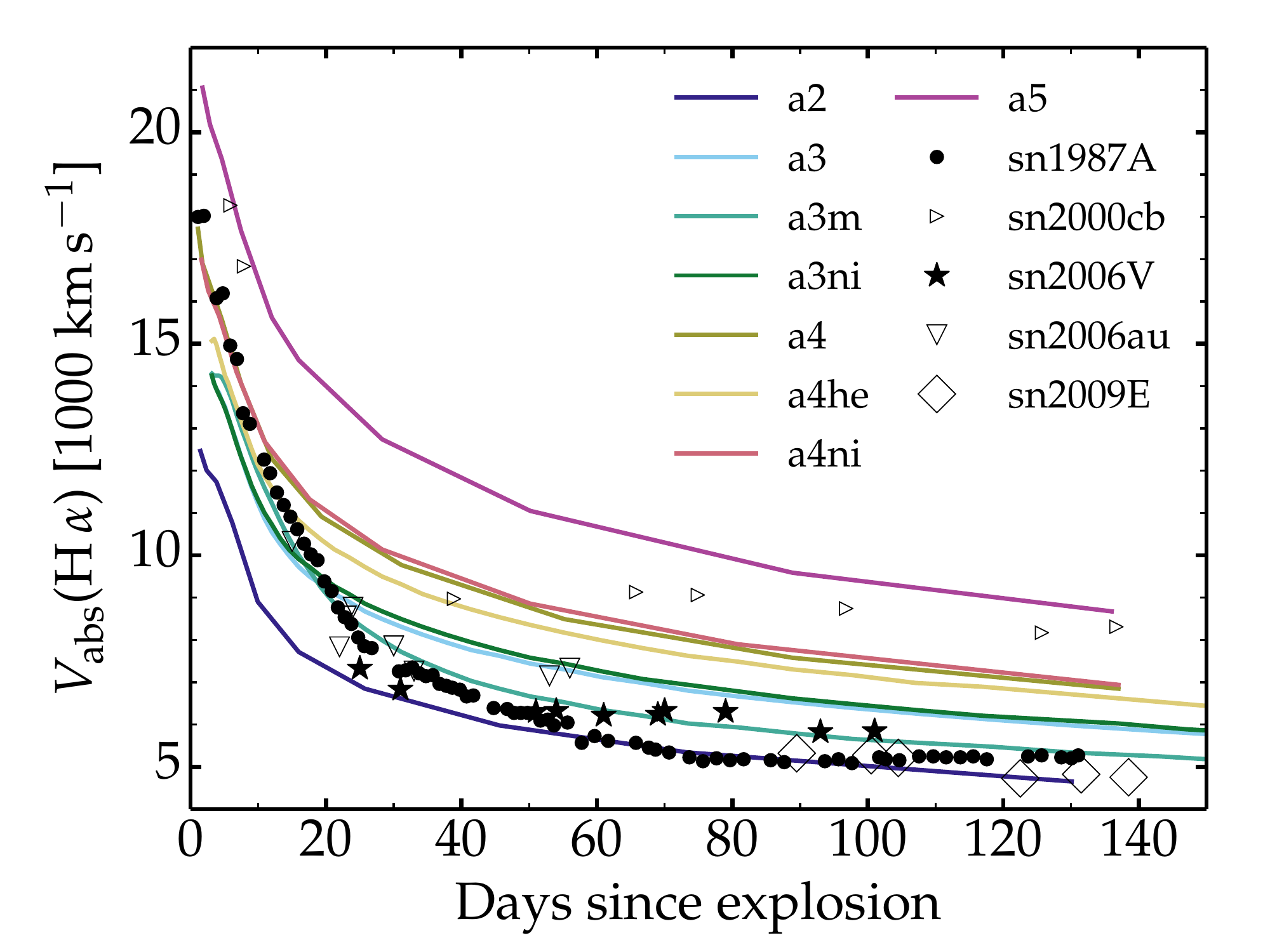}
   \includegraphics[width=0.97\hsize]{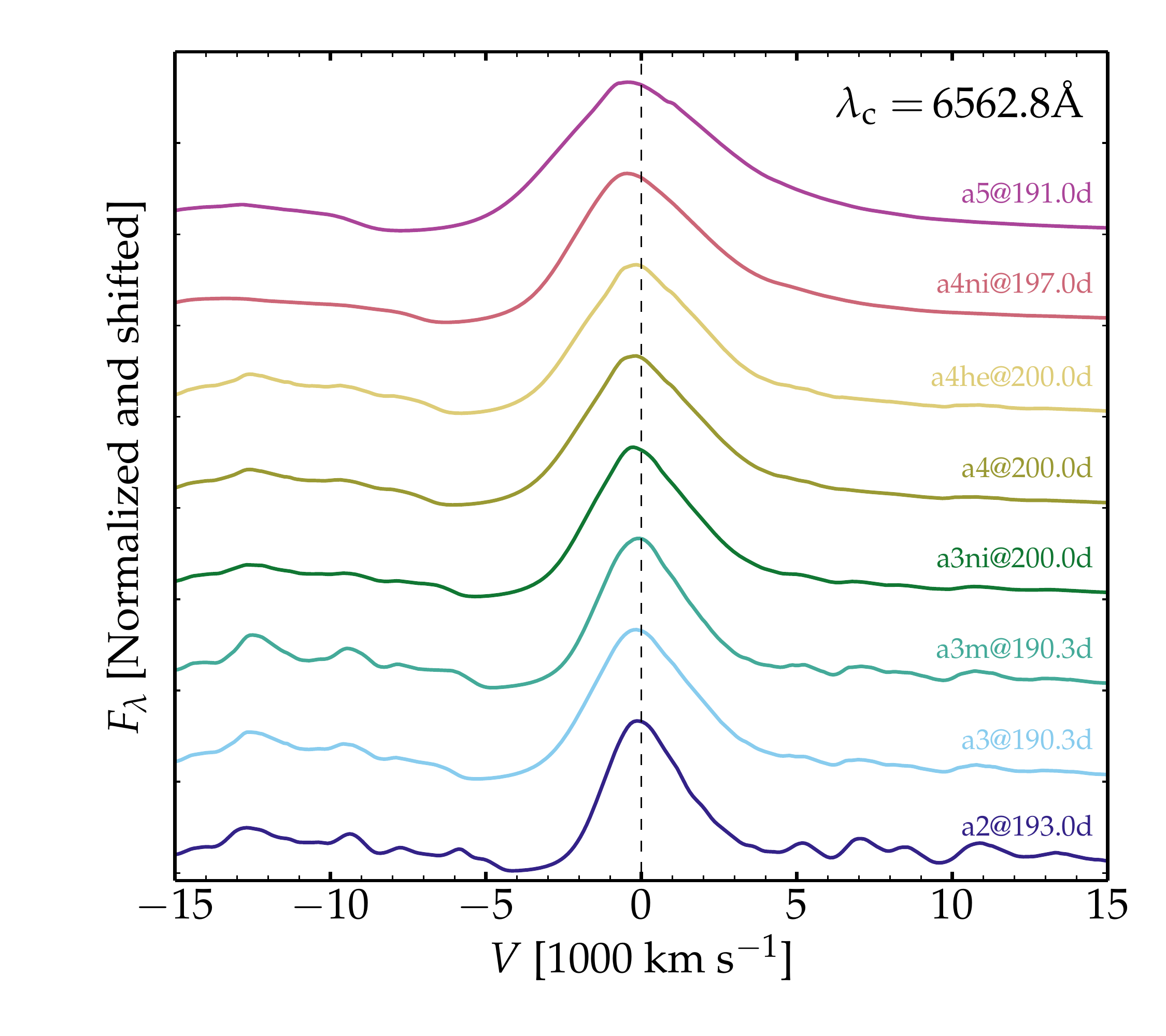}
\caption{Top: Doppler velocity at maximum absorption in Fe\two\,5169\,\AA\
for our set of models. We also overplot our measurements for Type II-pec SNe
1987A, 2000cb, 2006V, 2006au, and 2009E.
Middle: Same as top, but now for H$\alpha$. A gaussian smoothing with
a width of $10-20$\,\AA\ is used for noisy spectra or when the H$\alpha$
suffers from strong overlap (e.g., for SN\,2006au). In the latter case, the
measurement is merely indicative.
Bottom: Montage of spectra for our set of models showing the H$\alpha$ line
profile in Doppler-velocity space at about 200\,d after explosion.
\label{fig_v_ha}
}
\end{figure}

Figure~\ref{fig_color} shows the color evolution for our sample.
Despite the wide range in properties, all models follow a very similar
evolution, with only a slight offset between them.
Early on (i.e., say between 1 and $\sim$\,10\,d), the $U-V$ color
changes by four to five magnitudes as the photosphere cools down to the
H recombination temperature, and then it levels off at about 3.5\,mag
(Fig.~\ref{fig_phot_prop}).
Over that time, the change in $V-I$ (optical color) is only $0.5-1$\,mag,
while the change in $V-H$ (optical versus NIR brightness) is $1-2$\,mag.
Once the SN is well into the recombination phase (i.e., after about 20\,d),
the color is essentially constant until the ejecta becomes optically thin.
The color-color diagram shown in Fig.~\ref{fig_col_col} illustrates the
very similar evolution followed by all models, and suggests that such diagrams
may be used for reddening determinations.

The greater source of color contrast between models at early times is caused by mixing
(model a3m versus the rest).
At nebular times, the greater the \nifs\ mass, the bluer the optical color, although
\nifs\ mixing plays an important role (see also results for models YN1, YN2, and YN3
in \citealt{lisakov_08bk_17}).
Clumping is also a source of scatter in color \citep{d18_fcl}.

\begin{figure}
  \includegraphics[width=\hsize]{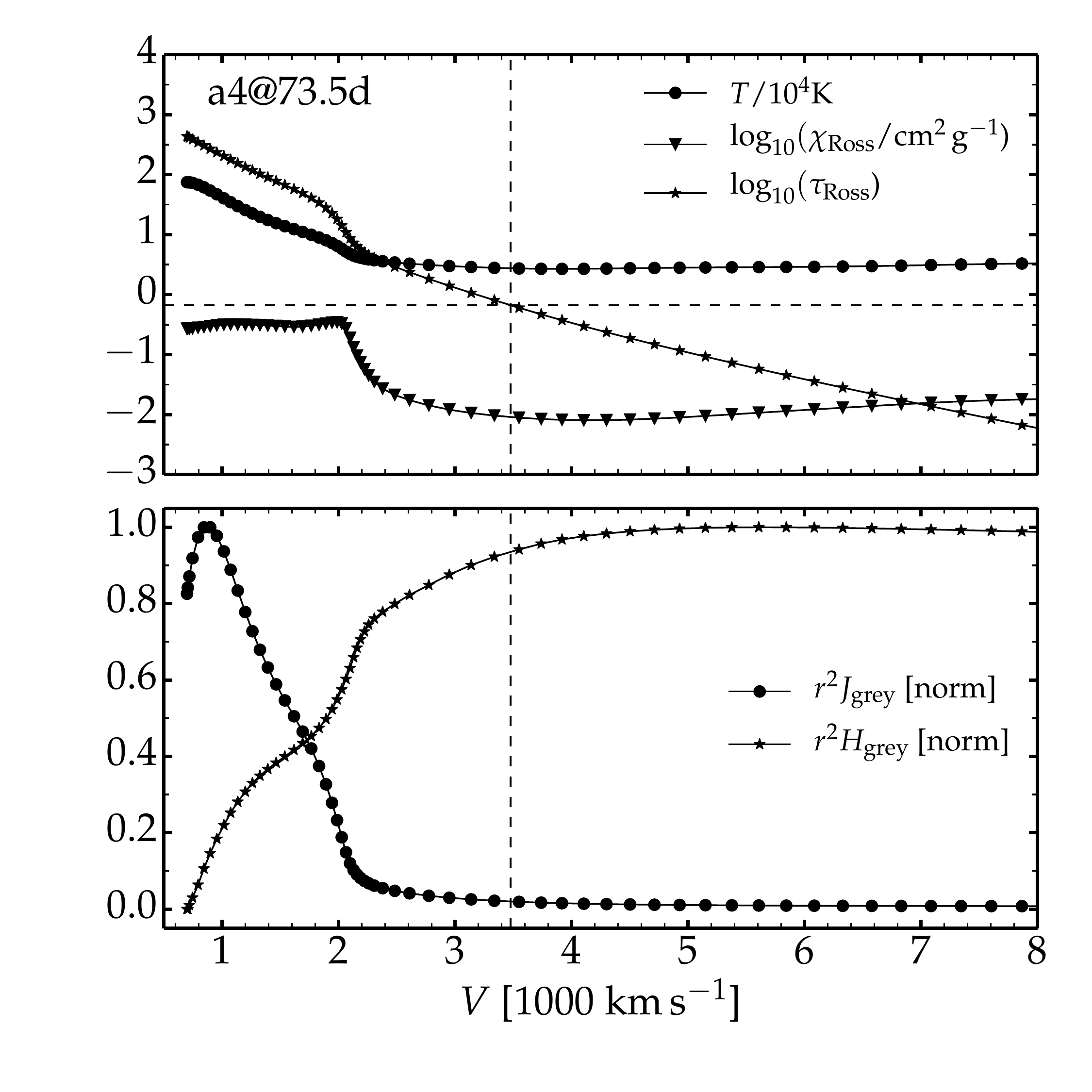}
\caption{Photospheric structure at 73.5\,d after explosion for model a4.
The top panel shows the temperature, the Rosseland-mean opacity,
and the Rosseland-mean optical depth. The dashed lines help visualize
the location of the photosphere (determined using the Rosseland-mean opacity).
The bottom panel shows the evolution of the gray mean intensity and the gray flux
(scaled by $r^2$, where $r$ is the local radius, and subsequently normalized)
computed by \cmfgen.
\label{fig_phot_struc}
}
\end{figure}

\section{Evolution of spectral properties}
\label{sect_spec}

Figure~\ref{fig_spec_3epochs} shows the spectra for our set of models
at 2.1, 25.7, and 97.8\,d after explosion. The first striking property is that
all model spectra look nearly identical (at a given epoch)
apart from the difference in line widths.
The H$\alpha$ line profile always reveals a clear P-Cygni profile with
very little structure (see also the bottom panel of Fig.~\ref{fig_v_ha}).
In all models (and thus for a range of \nifs\ mass and mixing),
He\one\,5875\,\AA\ is predicted for up to about 2\,d even with
a helium mass fraction in the outer ejecta of 0.27. Producing this line requires
a non-LTE treatment, as was already obtained and discussed in \citet{DH05a} --- in
contrast, \citet{eastman_87A_89} required an unrealistically large helium mass fraction
(of about 0.9) to reproduce this line.
Si\two\,6355\,\AA\ causes a blueshifted kink
in model a2 at 2.1\,d. In general, Ba\two\,6496.9\,\AA\ does not impact H$\alpha$,
even in the low-energy model a2.
In contrast, Ba\two\,6496.9\,\AA\ is very
strong in the model X of \citet{lisakov_08bk_17} for SN\,2008bk,
and indeed causes strong features on top of H$\alpha$ in all low-energy
SNe II-P \citep{roy_08in_11,lisakov_08bk_17,lisakov_ll2p_18}.
The multiplet line O\one\,7774\,\AA\ is present in all our models
after a few days (this line should not be mistaken at late times with the doublet
K\one\,7665--7699\,\AA).
During the recombination phase, the spectrum evolves little, except
for the progressive reddening of the SED and the narrowing of line profiles
(see middle and right panels of Fig.~\ref{fig_spec_3epochs}).

The H$\alpha$ line is strong and broad at all times in all our models, even
for low \nifs\ mass or mixing. A similar result was already obtained in \citet{DH10},
who argued that time dependence is key for reproducing this feature.
\citet{mitchell_56ni_87A_01} argued that \nifs\ mixing and the associated non-thermal
effects are key to reproduce Balmer lines as soon as 4\,d after explosion in SN\,1987A.
But in the simulations of \citet{DH10}, Balmer lines are strong for 20\,d (the time coverage
of the models) even though non-thermal effects were not treated at the time in \cmfgen.
In the present simulations as well as in \citet{DH10}, the influence of \nifs\ is negligible
both for the light curve and for the spectra at these early times.

The H$\alpha$ emission component, whose wavelength at maximum has
a significant blue shift early on, exhibits no obvious skewness (beyond that
expected for a P-Cygni profile).
The H$\alpha$ emission flux appears stronger than the absorption
part, and stronger than any other line in the optical spectrum.
These properties are expected for a spherically-symmetric smooth ejecta
in which the spectrum formation region recedes slowly and monotonically
through a monotonically increasing density profile.

Figure~\ref{fig_v_ha} shows the Doppler velocity at which the location
of maximum absorption occurs in H$\alpha$ and Fe\two\,5169\,\AA.
These measurements are, at best, indicative of the expansion rate
because different lines often show very different width, and often
the width of the absorption is very different from the width of the emission
(see Section~\ref{sect_spec_obs}).
It confirms what
Fig.~\ref{fig_spec_3epochs} illustrates in that each model shows a continuous
and monotonic evolution from large to small Doppler velocities.
For models that differ in ejecta kinetic energy, the trajectories do not cross.
In other words, an offset (of the same sign
and relative magnitude)  always remains between
models of different energy (here the ejecta mass is roughly the same in all models).
A weaker \nifs\ mixing leads to narrower H$\alpha$ and Fe\two\ lines for
the same kinetic energy (compare models a3 and a3m), but the offset
is only sizable at late times when non-thermal processes are strong at the photosphere.
The width of H$\alpha$ (whether in absorption or in emission) remains
large even at nebular times (thus even when a photosphere no longer exists) because
of the combined effects of ionization freeze-out and non-thermal effects
(which may strengthen in the outer ejecta as the $\gamma$-ray mean
free path increases). At nebular times, H$\alpha$ forms over a large volume of
the ejecta, spanning from the inner regions and going nearly all the way to the
maximum ejecta velocities.

Observations (overplotted in the top two panels of Fig.~\ref{fig_v_ha})
reveal a very chaotic
behavior. SN\,2000cb is the only SN in the sample for which the H$\alpha$
evolution is compatible with one model (i.e., a5).
While model a4 matches SN\,1987A early on, the two progressively diverge
and beyond 30\,d, it is the least energetic ejecta model a2 that corresponds
the most closely to SN\,1987A.
For SN\,2006V, the closest match is model a3 (intermediate ejecta energy between
models a2 and a4) while its brightness at maximum is the greatest of all SNe II-pec
in our sample, and greater than any of our models. This seems in conflict with
the notion that stronger Type II SN explosions generally produce a greater amount
of \nifs\ \citep{hamuy_03,sukhbold_ccsn_16,muller_56ni_17}.

As we discuss further below and in Section~\ref{sect_conc}, the
differences with our model predictions are probably indicative,
in part, of departures from spherical symmetry, the presence of clumping or
of chemical inhomogeneities, and perhaps of a diversity of power sources
at the origin of the SN luminosity. Variation in progenitor radius, mass,
and metallicity may also play a part.

\section{Photospheric structure at the recombination epoch}
\label{sect_phot_struct}

Figure~\ref{fig_phot_struc} shows the photospheric structure of model a4 at 73.5\,d
after explosion (hence during the recombination epoch and close to bolometric
maximum). This structure, which is characteristic of all our models at the recombination
epoch, shows that the location of the photosphere (given by the intersection of the
two dashed lines) is outside of the H-recombination front, which is where all curves
show a jump.  The jump in opacity is related to the jump in electron density
across the recombination front, and causes the jump in optical depth
at $\sim$\,2200\,\kms. This jump in optical depth is about 1000\,\kms\
deeper than the photosphere and causes the jump in mean intensity and flux.
The flux progressively increases from the front and is nearly maximum at the photosphere.

The radius or velocity offset between the location of the recombination front and
the location of the photosphere arises from the partial ionization of the gas
above the photosphere. This ionization freeze-out is caused by a time-dependent
effect \citep{UC05,D08_time}.

The temperature profile above the recombination front is very flat, and shows no jump across the photosphere. The Rosseland-mean opacity in these regions does not drop below 0.01\,cm$^2$\,g$^{-1}$ (in our simulations, this opacity even increases outwards because the temperature rises outwards, probably because of non-thermal (decay) heating at low density). In \cmfgen, the temperature (together with all level populations) results from balancing the cooling rates and the heating rates while requesting charge neutrality. The temperature structure in regions of optical depth less than about ten (i.e., from below the photosphere and beyond) is very different from the one that would result from the flux-limited-diffusion (FLD) approach often used in radiation hydrodynamics codes (which assume LTE for the gas, and infer the flux from the gradient of the mean intensity or the Planck function; see, e.g., \citealt{mm84}). Consequently, the FLD flux computed from such a temperature structure looks unphysical (irrespective of the choice of flux limiter). For example, the optically-thin regions where the non-LTE temperature profile is flat would give a zero flux in FLD while the (total) flux in \cmfgen\ drops as $1/r^2$ in spite of this temperature structure (see Fig.~\ref{fig_phot_struc}). In FLD, the temperature must decrease through the photosphere and above in order to carry the flux to infinity. Although this is expected, this emphasizes the fact that radiation-hydrodynamics codes calculate the emergent flux through a completely different physical reasoning and a completely different set of equations compared to a non-LTE time-dependent radiative transfer code like \cmfgen.


\section{Comparison to observations of Type II-pec SNe}
\label{sect_spec_obs}

   In this section, we compare the $V$-band light curves and
multiepoch spectra for the selected sample
of observed Type II-pec SNe with results from our model set.
For the observations, we use the SN characteristics described
in Section~\ref{sect_obs} and summarized in Table~\ref{tab_obs}.
Our model set was not designed to reproduce these observations
(one obvious offset is seen at nebular times when the decay power for
our adopted \nifs\ mass can be offset from the observed luminosity).
So, what we present here is a comparison, using the models
to guide our understanding of what may be the origin of these events.
We start with observed SNe for which we have models that provide
a fair match to the light curve or the spectra. We then turn to SNe II-pec
whose properties are harder to reproduce with a BSG star explosion model
powered by \nifs\ decay.

\begin{figure}
   \includegraphics[width=\hsize]{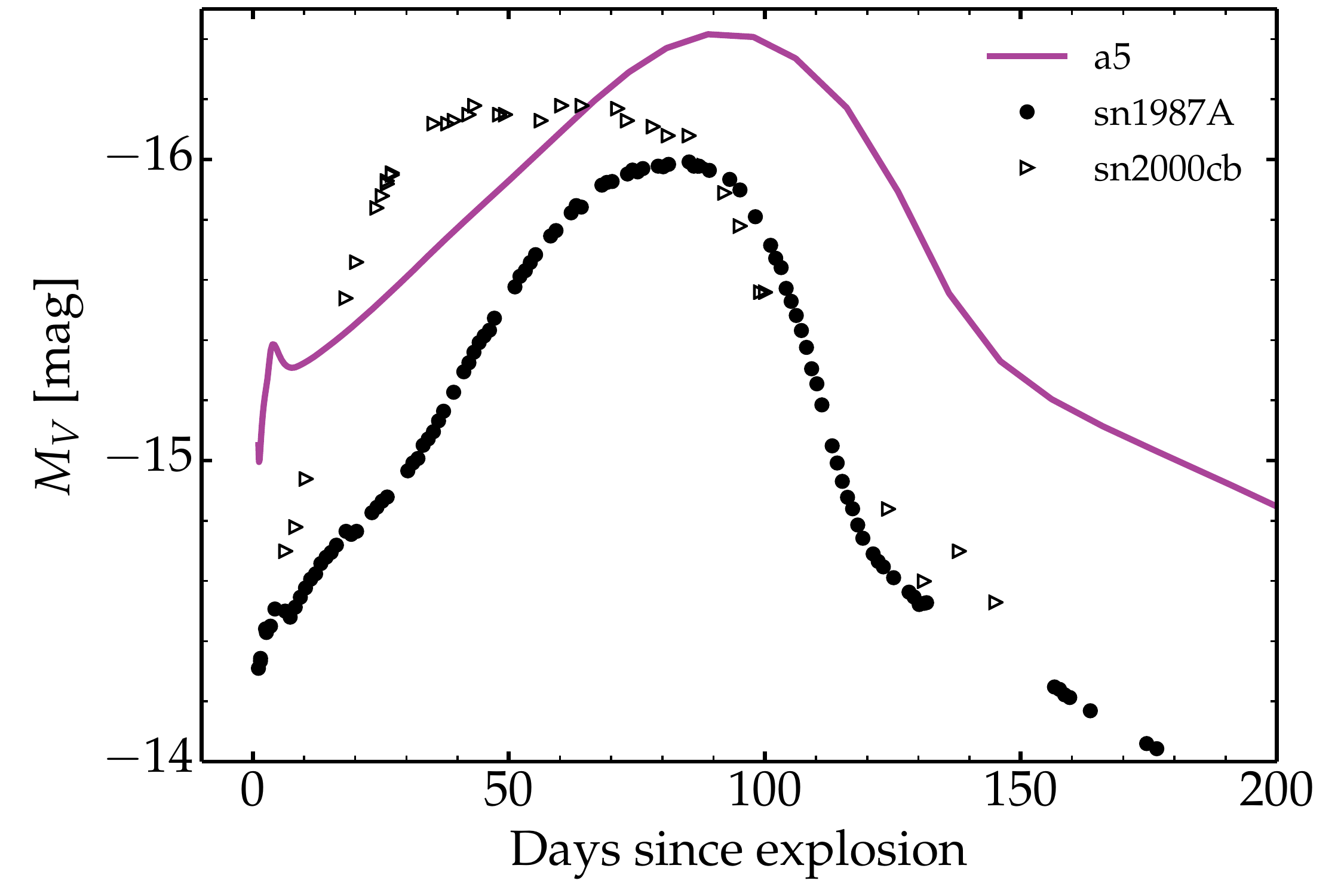}
   \includegraphics[width=\hsize]{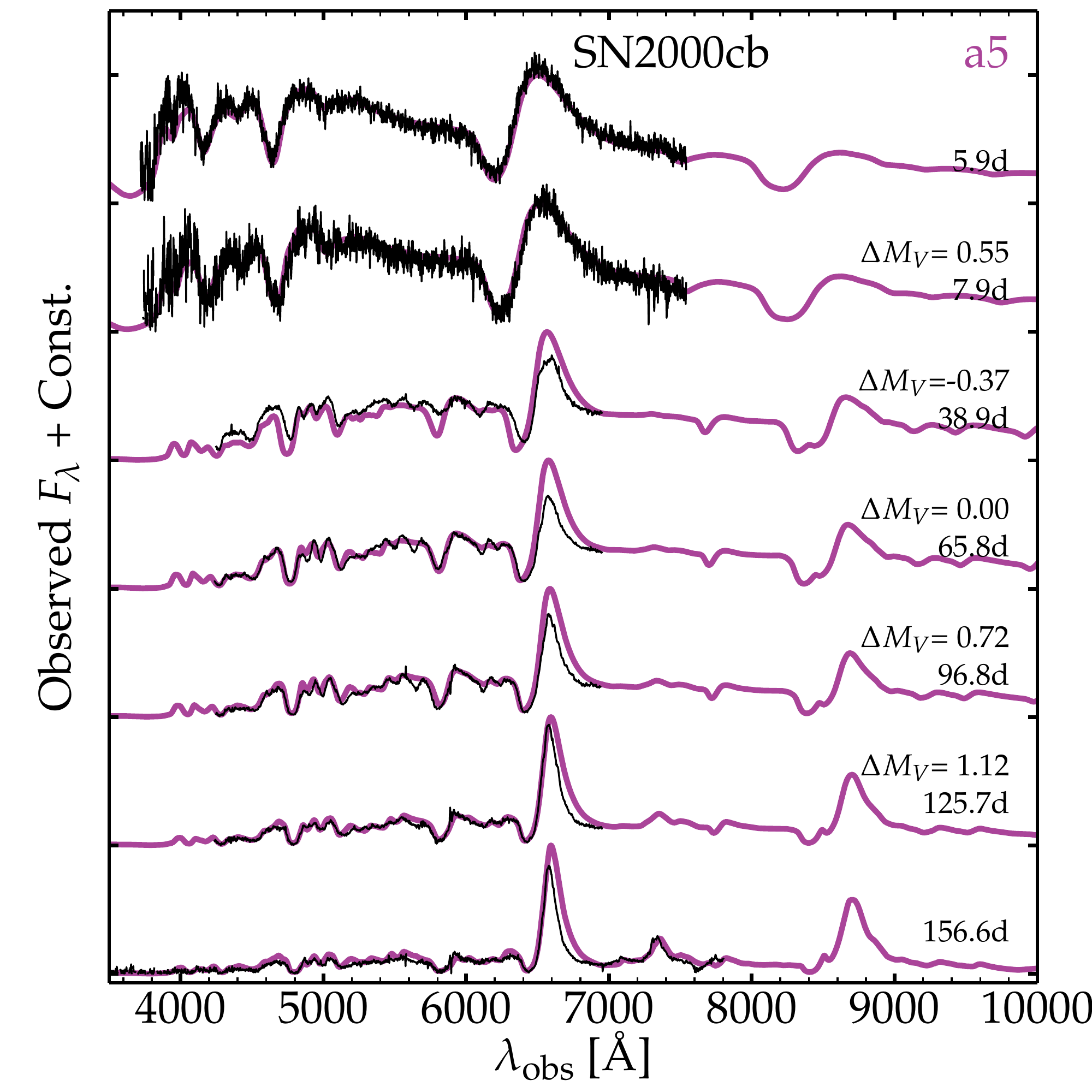}
\caption{Top: Absolute $V$-band light curves for SN\,2000cb
together with the results for model a5 (we also add SN\,1987A).
Bottom:  Multiepoch spectra for models a5 compared with the observed
spectra of SN\,2000cb. The model is redshifted and reddened (see Section~\ref{sect_obs}).
At each epoch, the spectra are normalized (in this case at 6000\,\AA) and then shifted vertically.
The label $\Delta M_V$ gives the $V$-band magnitude difference between observations and model
at each time (whenever we can interpolate between photometric data points).
\label{fig_a5_00cb}
}
\end{figure}

\subsection{Comparison to SN\,2000cb}
\label{sect_spec_00cb}

Figure~\ref{fig_a5_00cb} compares the $V$-band light curve (with respect
to the time of explosion) and the multiepoch spectra of SN\,2000cb
with the higher energy explosion model a5.
The $V$-band light curve of SN\,2000cb differs from that of
the prototypical Type II-pec SN\,1987A (see top panel of Fig.~\ref{fig_a5_00cb}).
Starting at $-14.6$\,mag, it brightens
to maximum in only 25 days (this is a bolometric rise, not a color effect),
which is more typical of Type I SNe \citep{conley_snia_rise_06,drout_11_ibc}.
It then stays at the same $V$-band brightness of about $-16.2$\,mag for about 60\,d
before declining abruptly and becoming nebular at about 110\,d after explosion.
Photometrically, our model a5 departs sizably from the behavior observed for SN\,2000cb.
Model a5 is 0.5\,mag brighter at the earliest times, then rises slowly to a maximum of
$-16.4$\,mag (0.2\,mag brighter than SN\,2000cb) at 90\,d. The transition to the nebular
phase occurs on a similar time scale with the same drop in magnitude. Model a5 is
overluminous at nebular times by about 0.5\,mag (the \nifs\ mass is too high
by about 60\%).

The offset in brightness is therefore at the level of 0.5\,mag at most times,
which translates into an offset in luminosity of 60\%, or 25\% in
photospheric radius. Hence, this discrepancy is not so large. Reducing
the \nifs\ mass to 0.1\,\msun\ would probably resolve the discrepancy
at and beyond maximum. At earlier times, asymmetry might explain
the very fast rise (see \citealt{utrobin_chugai_00cb_11}; see below).

Spectroscopically, the agreement is generally good at all times shown (from 4.9 until 155.6\,d after explosion) with only a few discrepancies. The worst match is seen at 38.9\,d, when the model overestimates the strength and width of H$\alpha$, H$\beta$, and Na\one\,D.

A persistent discrepancy is seen for the H$\alpha$ profile, which, as discussed
in Section~\ref{sect_spec}, shows a smooth evolution in all models
(i.e., the line narrows in time and remains strong both in absorption and emission).
In SN\,2000cb, H$\alpha$ is initially broad, but then appears
narrow and weak (i.e., narrower and weaker than in model a5). Subsequently,
the line remains broad and strengthens relative to the continuum until the last epoch shown.
Apart from the first two epochs, the H$\alpha$ emission is much
broader in the model than in the observations, although the discrepancy
weakens at later times. The agreement tends to be better for the absorption
component than for the emission component (both in strength and width) -- the absorption samples
material along the line of sight to the SN, while the emission component arises from a larger
volume, including lines of sight that do not intersect the SN photosphere.
These H$\alpha$ discrepancies may be attributed
to clumping \citep{d18_fcl}, or asymmetry \citep{utrobin_chugai_00cb_11},
or both.

H$\beta$ is always present (unlike in SN\,1987A or SN\,2009E, in which
it is absent for many weeks at the recombination epoch;
see Sections~\ref{sect_spec_87A} and \ref{sect_spec_09E}).
The discrepancies that affect H$\alpha$ do not seem to affect H$\beta$,
which is well reproduced by the model at all times except at 38.9\,d.
The same agreement holds for Na\one\,D (absent at the first two epochs,
poorly matched at 38.9\,d, well matched at later times).

These various levels of agreement highlight one major problem in the inference
of the expansion rate (and explosion energy) of SN ejecta. Furthermore, a line
may be well fitted in absorption but poorly in emission. There is some
arbitrariness in relying on either. A better approach is probably to
assess the overall match to the observed spectrum rather than arbitrarily
adopting one line.

The red part of the optical range is lacking except in the last observation of SN\,2000cb,
At that time, the Ca\two\,7300\,\AA\ is observed and model a5 matches
the strength  well (as we will see below, this is not always the case, as
for SN\,2009E; Section~\ref{sect_spec_09E}).

Overall, a better model for this SN may be obtained by adopting
a lower \nifs\ mass of about 0.1\,\msun, which would probably resolve
the discrepancy in the light curve at and beyond maximum.
At early times, the faster rise to maximum and the rapid narrowing
of the H$\alpha$ line probably requires asymmetry,
as proposed by \citet{utrobin_chugai_00cb_11}.
In this context, the explosion energy of SN\,2000cb is probably
lower than in model a5 ($2.46 \times 10^{51}$\,erg), and may even
be standard for a core-collapse SN since the bulk of the mass, located at
low velocity, would then carry little kinetic energy.
The same effect likely impacts the inference
for GRB/SNe like 1998bw \citep{dessart_98bw_17}.
The ejecta mass of model a5 is 13.10\,\msun, much smaller
than the value of 22.3\,\msun\ proposed by \citet{utrobin_chugai_00cb_11}.
Reducing the \nifs\ mass by a factor of two would probably narrow the light
curve by 30\,d (like in the models a3ni and a3). The rapid rise in model
a5 seems hard to reconcile simply with strong mixing, since we already employ
strong mixing. It would also require more \nifs\ ejected in our direction.

The $V$-band light curve of SN\,2000cb is reminiscent of the morphology
obtained for magnetar-powered Type II SNe \citep{dessart_pm_18},
although it would require a more slowly-rotating magnetar than employed
in that study. Further work is needed to understand this object and resolve the
various discrepancies.

\begin{figure}
   \includegraphics[width=\hsize]{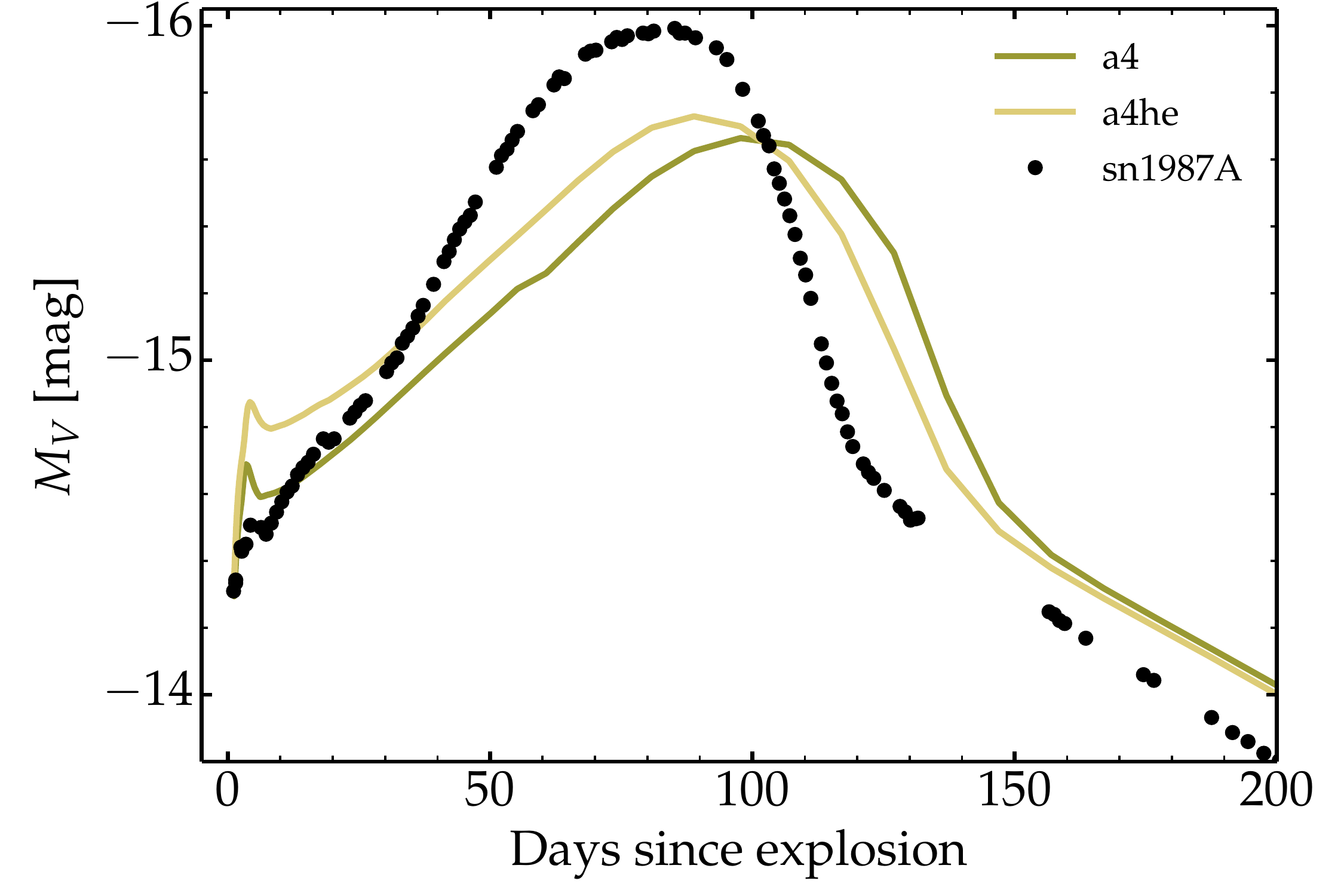}
   \includegraphics[width=\hsize]{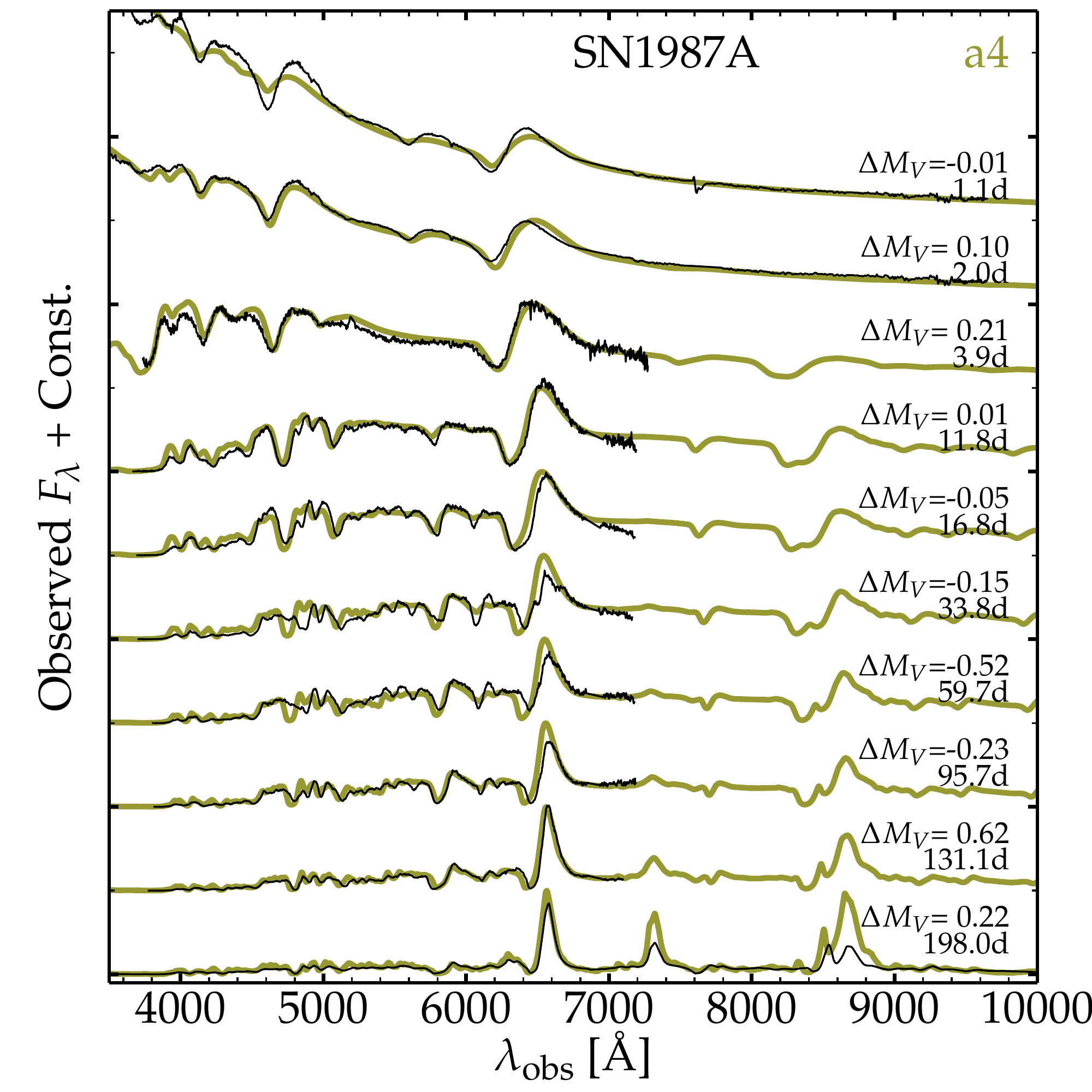}
\caption{Same as Fig.~\ref{fig_a5_00cb}, but now for SN\,1987A
and models a4 and a4he.
\label{fig_a4_87A}
}
\end{figure}

\subsection{Comparison for SN\,1987A}
\label{sect_spec_87A}

SN\,1987A has been and is still extensively studied due to its proximity.
Its progenitor was identified as a BSG progenitor star, which is also supported by
its long and slow rising light curve to an optical maximum at about 90\,d after
explosion. Numerous independent observations suggest that the explosion
was asymmetric on both small and large scales.
From late time observations,
evidence that the inner ejecta is clympy and asymmetric comes from
radio imaging of molecular emission lines \citep{abellan_87A_17},
from standard spectroscopy \citep{fransson_chevalier_89, spyromilio_87a_90,li_87A_93,jerkstand_87a_11,jerkstrand_04et_12},
and from integral field spectroscopy \citep{kjaer_87a_10}.
Asymmetry is also suggested from early time observations by
the fine-structure in the H$\alpha$ line profile \citep{hanuschik_87a_88},
the early-time detection of intrinsic polarization \citep{hoeflich_87A_91,jeffery_87_pol_91},
the direction-dependent spectra of SN\,1987A observed through light echoes \citep{sinnott_87a_13},
or the smooth rising optical brightness and the high-energy radiation observed after about 200\,d
(see, e.g., \citealt{sn1987A_rev_90}). Despite this evidence, all the radiative-transfer modeling
for SN\,1987A has been limited to 1-D. Multi-D effects can in some cases be taken into
account in a simplistic manner. For example, chemical mixing can be treated in a 1-D model,
as done here. We can also investigate the effect of clumping in 1-D, as shown in
\citet{d18_fcl}. Although inadequate, the assumption of spherical symmetry
can help identify the signatures of asymmetry in light curves and spectra.

With this in mind,  Fig.~\ref{fig_a4_87A} compares the $V$-band light curve (with respect to the time of explosion) and the multiepoch spectra of SN\,1987A with model a4 (model a4he is also shown in the top panel). The $V$-band light curve of model a4 rises slowly from $-14.3$\,mag at 1\,d to a maximum at 100\,d before turning nebular at 150\,d. The evolution of SN\,1987A differs. Although it has the same brightness at 1\,d as model a4, it rises in only 80\,d to a maximum 0.4\,mag brighter and turns nebular at 120\,d. Model a4he, with its greater He to H abundance ratio, has roughly the same brightening rate as model a4 on the way to maximum, peaks at the same time as SN\,1987A but at a fainter magnitude by 0.3\,mag. Overall, the light curve morphology of these models is analogous to that of SN\,1987A, but with a slower rise to a fainter maximum. Our light curves are overall too broad.

The bottom panel of Fig.~\ref{fig_a4_87A} shows multiepoch spectra for SN\,1987A
and for model a4. The agreement is fair at early and late times, but there are discrepancies
from about one month until the time of maximum.
In the observations at 33.8\,d, SN\,1987A shows a narrower H$\alpha$ profile,
strong Ba\two\ lines, and no H$\beta$ line.  The structure in H$\alpha$
is reminiscent of what is observed in low-luminosity Type II-P SNe
\citep{roy_08in_11,lisakov_ll2p_18}. In the case of SN\,2008bk,
\cmfgen\ predicts that the structure in H$\alpha$ is caused by overlap
with Ba\two\,6496.9\,\AA, and that this feature is generally not seen in
standard-energy SNe II-P because the lines are broader
(for the same reason, the low-energy Type II-pec SN \,2009E also
shows strong Ba\two\ lines; see Section~\ref{sect_spec_09E}
and \citealt{pasto_9e_2pec_12}).  In \citet{d18_fcl},
we show that clumping can help reduce the ionization of the gas and
enhance the abundance of Ba$^+$. This can boost the Ba\two\ line strength and
may cause  the peculiar morphology  of H$\alpha$ and the absence of H$\beta$
(which overlaps with Ba\two\,4899\,\AA\ and Ba\two\,4934\,\AA).
\citet{UC05} proposed that the strong Ba\two\, lines were the results of
a time-dependent effect. But we include time dependence in our simulations
and do not predict strong Ba\two\ lines here. Increasing the Ba abundance
by a factor of five at 35\,d does not resolve the problem. It is unlikely that
the strong Ba\two\ lines are caused by a generic process like time dependence
because this feature is not always seen. In our sample of SNe II-pec,
SNe 1987A and 2009E are the only objects that show strong Ba\two\ lines.

Figure~\ref{fig_v_ha} shows that the Doppler velocity at maximum absorption
in H$\alpha$ and Fe\two\,5169\,\AA\  initially follows the predictions for
model a4 but quickly transitions to resembling the properties for the weakest
explosions in our sample. Reducing the \nifs\ mixing in our simulations
would produce narrower lines, but it would delay the rebrightening
(compare model a3 and a3m, or see, e.g., \citealt{blinnikov_87A_00}).
In \citet{d18_fcl}, we argue that ejecta clumping can speed up
the recession of the photosphere, boosting the luminosity (and the brightening
rate) while at the same time producing narrower line profiles. \nifs\ mixing
cannot achieve this because it tends to produce broader lines, not narrower
lines.
Such clumping is fundamentally associated with chemical inhomogeneities
and implies ejecta asymmetry. It is unclear whether the explosion energy
is also asymmetric (i.e., if the shock-deposited energy varies with angle).

\begin{figure}[h!]
   \includegraphics[width=\hsize]{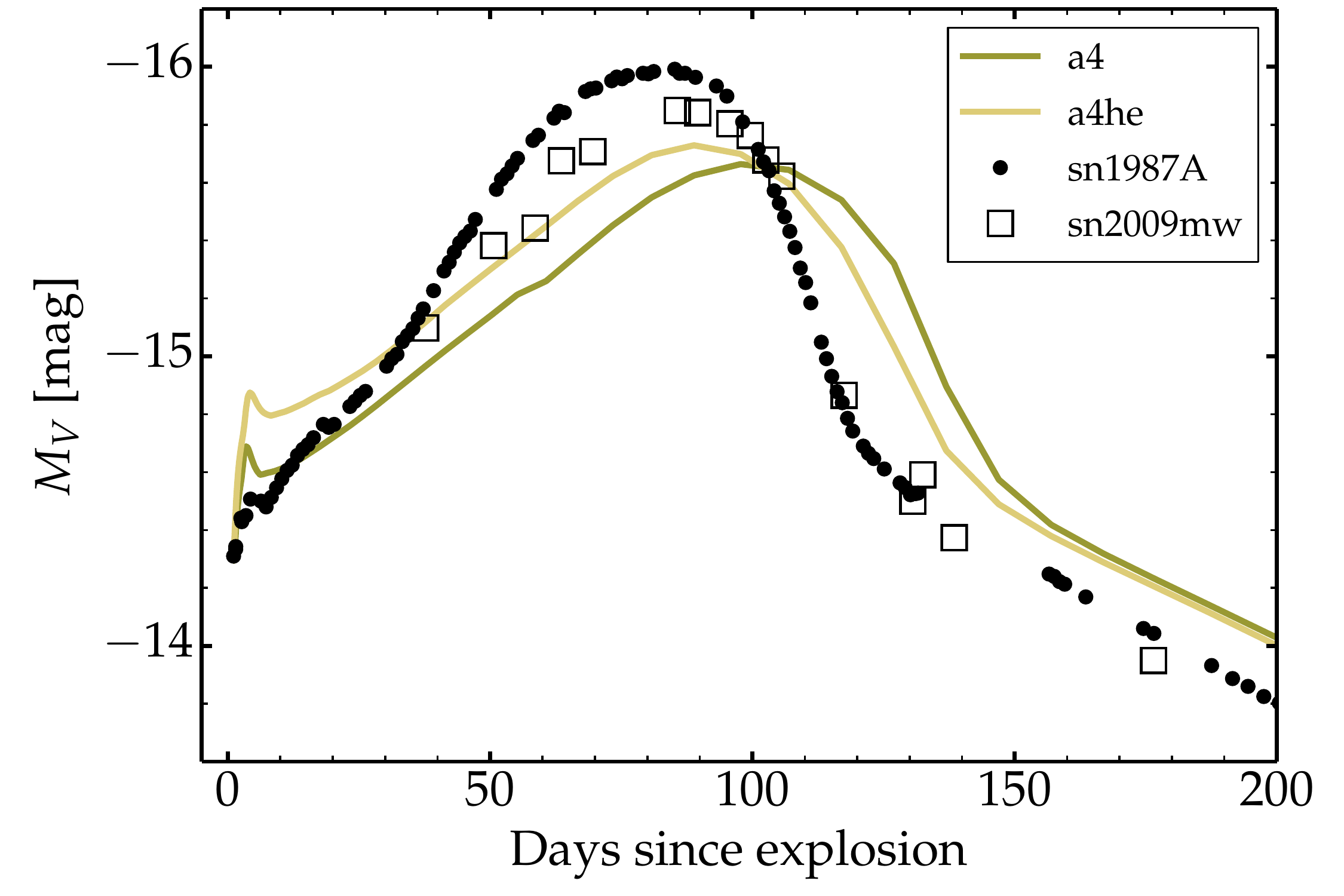}
   \includegraphics[width=\hsize]{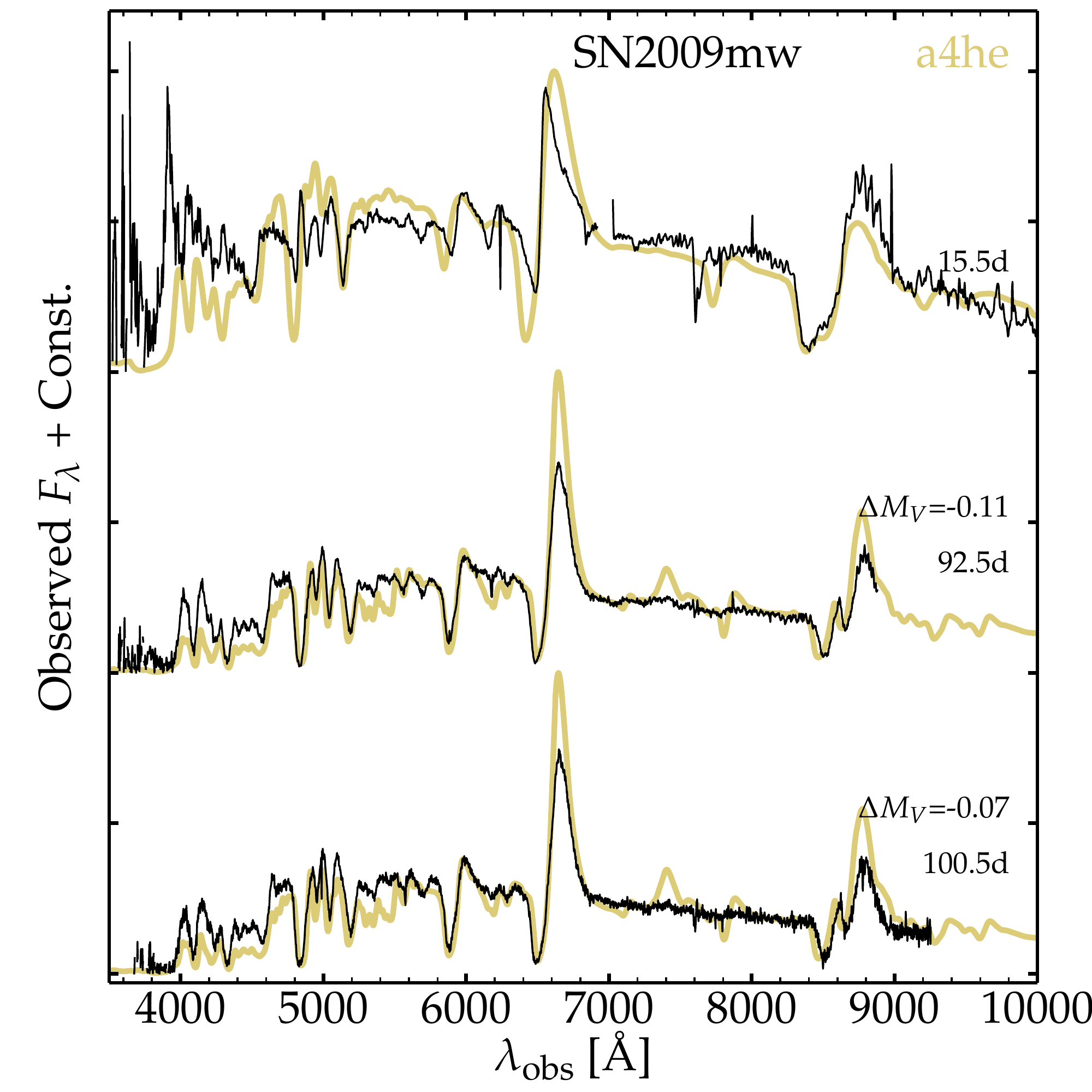}
\caption{Multiepoch spectra for models a4he compared with the observed
spectra of SN\,2009mw  \citet{takats_09mw_16}. The line profile and color
evolutions depart from expectations for a Type II-pec SN: lines are narrower early on;
there is no sign of Ca\two\,7300\,\AA\ nor O\one\,7774\,\AA; the optical colors are
bluer for longer. These features are unlikely to be exclusively a metallicity effect.
Some of these features are seen in iPTF14hls \citep{arcavi_iptf14hls}.
\label{fig_a4_09mw}
}
\end{figure}

Further work is needed to understand the mismatch in light curve properties,
in particular the brightening rate to maximum.
For the given ejecta mass and explosion energy, a combination
of clumping, weaker \nifs\ mixing, and a greater helium to hydrogen abundance
ratio will probably lead to a satisfactory match with \cmfgen.
Alternatively, a smaller mass would result in a faster rise to maximum and a brighter peak.
Simulations with other codes have yielded a good match to SN\,1987A
(with parameters that are close to ours; \citealt{woosley_87a_88},
\citealt{blinnikov_87A_00}; \citealt{utrobin_etal_15}), but these use
completely different techniques and make very little use of
spectral constraints.

\subsection{Spectral comparison for SN\,2009mw}
\label{sect_spec_09mw}

Figure~\ref{fig_a4_09mw} compares the $V$-band light curve  and the multiepoch spectra of SN\,2009mw with model a4he (SN\,1987A and model a4 are also shown in the top panel). Although the explosion date is uncertain, SN\,2009mw may have had a similar $V$-band light curve to SN\,1987A, with a slower brightening rate to a slightly fainter maximum \citep{takats_09mw_16}.

Our model a4he has a similar light curve to SN\,2009mw, with only slight offsets at the 0.2\,mag level (apart from the fall-off from maximum, which occurs about 10\,d later in the model -- this discrepancy is function of the adopted time of explosion). Spectroscopically, model a4he yields a satisfactory match for the last two epochs shown in Fig.~\ref{fig_a4_09mw}, with perhaps a small offset in color (the model is too red). The first spectrum may be inaccurate in relative flux. The H$\alpha$ profile is not well matched at 15.5\,d but is well matched for the last two epochs around the time of maximum. A peculiarity of this SN is the lack of O\one\,7774\,\AA\ and the Ca\two\ doublet at 7300\,\AA. O\one\,7774\,\AA\ is seen in SN\,2006au but not in SN\,2006V. The spectral range of our observations for SN\,1987A extends to 7200\,\AA\ in the optical so no information on O\one\,7774\,\AA\ is accessible for SN\,1987A (the spectra of \citealt{spyromilio_87A_91} suggest that O\one\,7774\,\AA\ is also very weak in SN\,1987A at the corresponding epochs). Of all SNe II-pec discussed in this paper, SN\,2009mw seems the closest analog of SN\,1987A, although numerous spectral properties differ (see also discussion in \citealt{takats_09mw_16}).

\subsection{Comparison for SN\,2009E}
\label{sect_spec_09E}

\begin{figure}[h!]
   \includegraphics[width=\hsize]{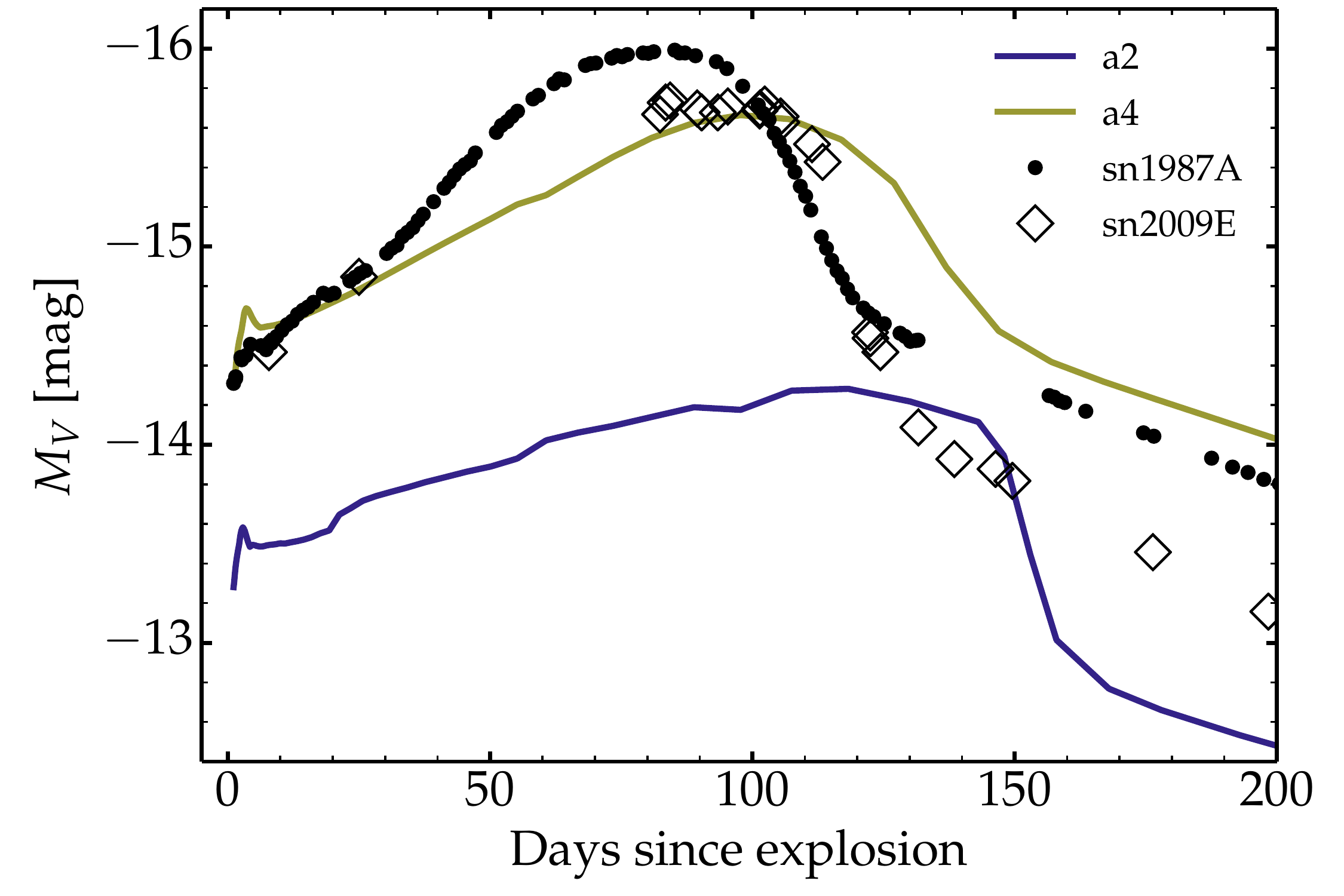}
   \includegraphics[width=\hsize]{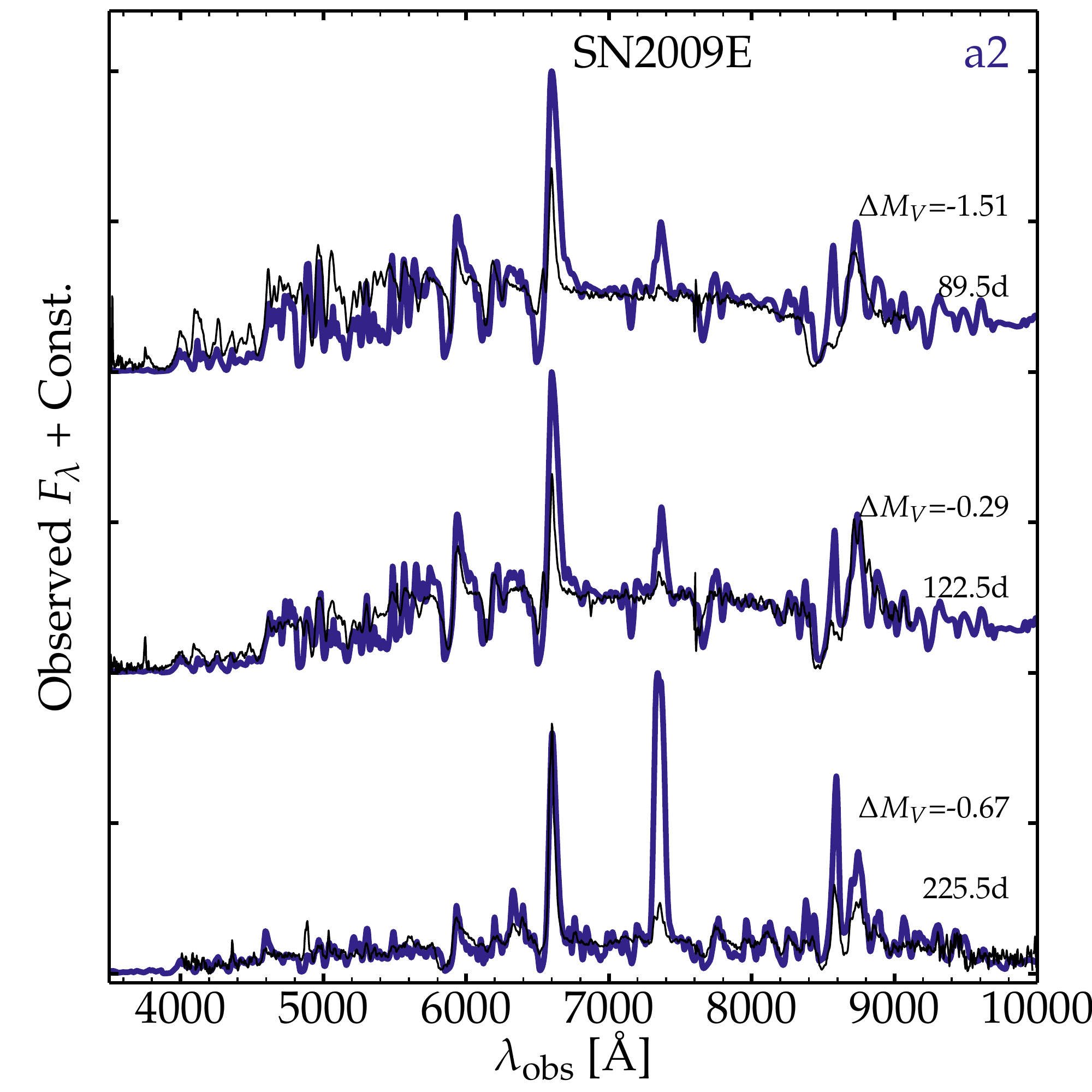}
\caption{Same as Fig.~\ref{fig_a5_00cb}, but now for SN\,2009E
and model a2 (the top panel also includes SN\,1987A and model a4).
\label{fig_a2_09E}
}
\end{figure}

  There is no good model in our set for SN\,2009E. \citet{pasto_9e_2pec_12}
presents SN\,2009E as a faint clone of SN\,1987A but
the $V$-band light curves of these two SNe are in fact very similar
(top of Fig.~\ref{fig_a2_09E}).
With a reddening $E(B-V)$ of 0.04\,mag, the $V$-band brightness is the same
for both at early times, but SN\,2009E brightens more slowly (at a rate
similar to the rate we obtain in all our models), transitions to the nebular phase
about 10\,d later. Its nebular-phase brightness suggests it has half the \nifs\
mass of SN\,1987A, hence about 0.04\,\msun\
(Fig.~\ref{fig_a2_09E}; \citealt{pasto_9e_2pec_12}).
Its decline rate at nebular times is faster, which may indicate
$\gamma$-ray escape in a lower mass ejecta (unlikely given the narrow lines,
which indicate a low expansion rate and a dense inner ejecta). However, this is
not a bolometric luminosity plot so this offset may be related to a different evolution
of the bolometric correction (the spectra for SNe 2009E and 1987A are
evidently different).

\begin{figure}
   \includegraphics[width=\hsize]{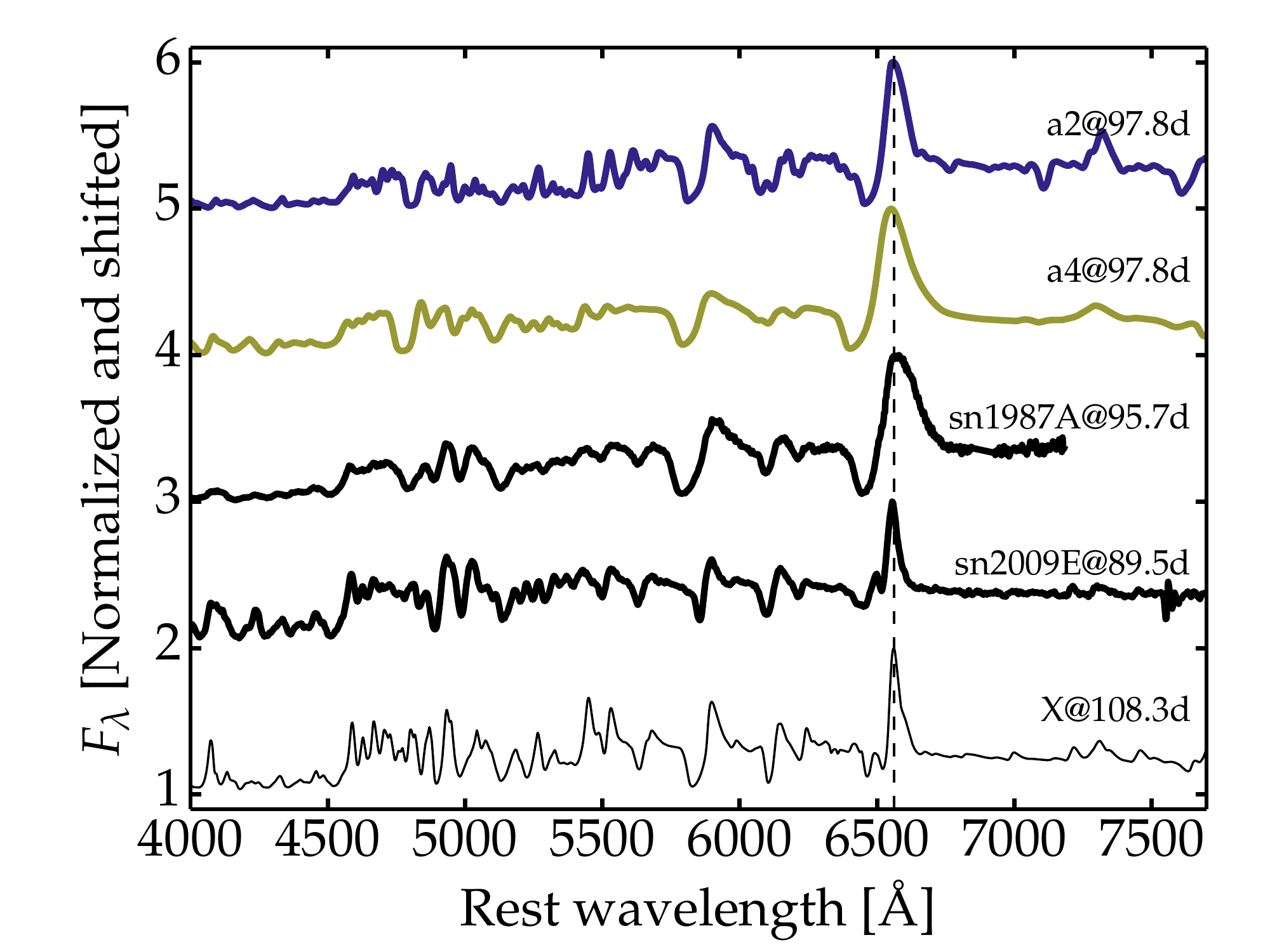}
\caption{Comparison between the spectra of SN\,2009E, SN\,1987A,
and models a2 and a4 around the time of maximum brightness.
We also overplot model X (used to model the low energy Type II-P SN\,2008bk;
\citealt{lisakov_08bk_17}) to show that a lower energy than used
for model a2 is needed to explain the very narrow H$\alpha$.
All spectra are normalized and shifted vertically for visibility. Observations
were corrected for redshift but not for reddening.
\label{fig_09E_08bk}
}
\end{figure}

Our model a4 (with ejecta kinetic energy of $1.24 \times 10^{51}$\,erg)
matches closely the $V$-band light curve of SN\,2009E,
with the exception of the larger brightness in the nebular phase.
But model a4 is much too energetic for SN\,2009E. Even our least
energetic model a2 (with ejecta kinetic energy of $0.47 \times 10^{51}$\,erg)
overestimates the width of the lines, although
the disagreement depends on the choice of line, what part of the
line is used, and what \nifs\ mixing is adopted (Fig.~\ref{fig_09E_08bk}).

The model X of \citet{lisakov_08bk_17}, which matches the SN\,2008bk spectral
evolution, matches closely the spectrum of SN\,2009E at 89.5\,d (Fig.~\ref{fig_09E_08bk}).
SN\,2009E is thus most likely a low energy explosion, with an expansion rate in
the inner ejecta that is much smaller than for SN\,1987A. There is no information at early
times to constrain how fast the outer ejecta was moving.
This spectral comparison also highlights the difficulty of inferring the explosion
energy from line profile widths.
In Fig.~\ref{fig_09E_08bk}, the Ba\two\ line at 6141.7\,\AA\
or the Fe\two\ line at 5169\,\AA\ has a similar appearance in SNe 2009E
and 1987A, as well as model X,
while the H$\alpha$ line profile of SN\,2009E is similar to that of model
X but considerably narrower and weaker than in SN\,1987A.
The factor two difference in \nifs\ mass between SNe 2009E and 1987A cannot
explain the large difference in H$\alpha$ line strength at bolometric maximum:
non-thermal effects for the formation of H$\alpha$ should be
efficient in both. Another peculiarity is that Na\one\,D appears broad in the
low energy model X but much narrower in SN\,2009E.
At 225\,d after explosion, model a2 overestimates the strength of the
Ca\two\,7300\,\AA\ doublet. This line, which is favored at lower densities relative
to the Ca\two\ NIR triplet, is much weaker in SN\,2009E than in SN\,1987A.

Overall, SN\,2009E is a very peculiar SN with a significant amount of
\nifs\ (if we assume this is the power source),
which allows it to rival the peak brightness of SN\,1987A, but with
a much lower explosion energy, probably on the order of that for SN\,2008bk
(see also \citealt{pasto_9e_2pec_12}).
The binding energy of a BSG star is quite large so a weak explosion should
lead to strong fallback, preventing the ejection of \nifs. One possibility is
that the BSG star progenitor is of moderate mass and binding energy, which would
cause limited fallback even for a weak explosion.
An alternative is that this SN is not powered by \nifs\ decay, but perhaps by
the compact remnant. The sparse photometric and spectroscopic coverage
of SN\,2009E does not help resolving this puzzle.

\subsection{Spectral comparison for SN\,2006V}

\begin{figure}
   \includegraphics[width=\hsize]{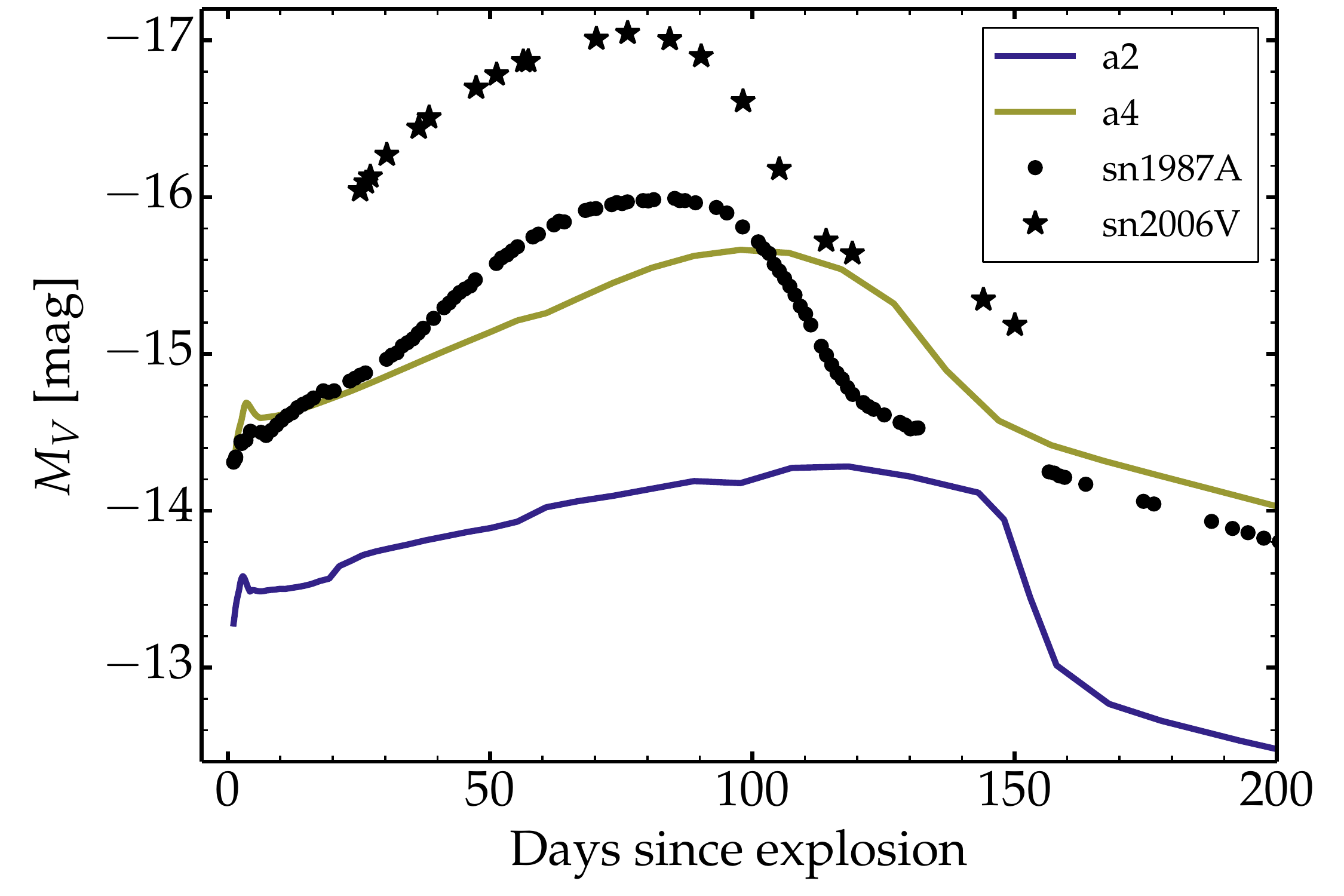}
   \includegraphics[width=\hsize]{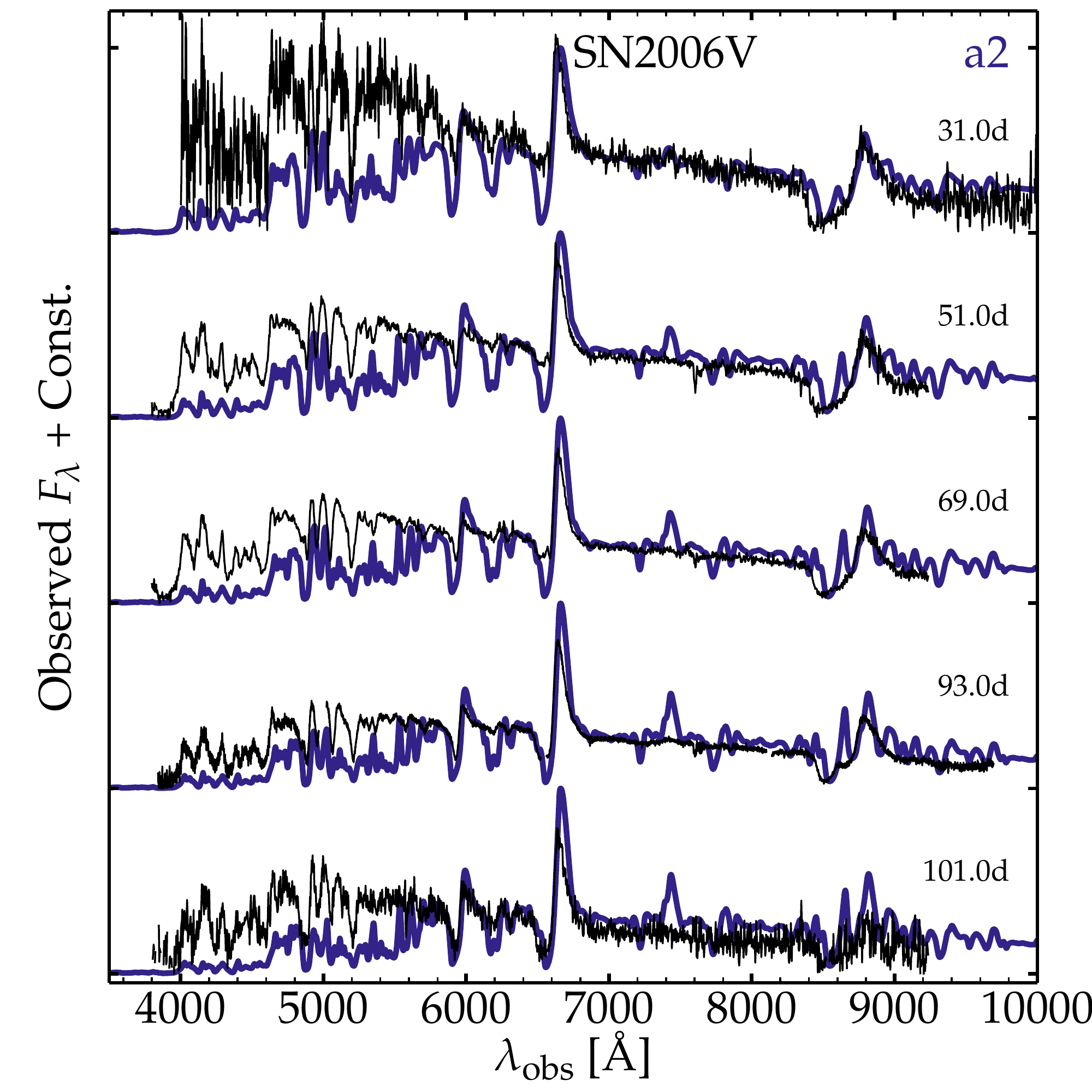}
\caption{Same as Fig.~\ref{fig_a5_00cb}, but now for SN\,2006V
and model a2 (the top panel also includes model a4 and SN\,1987A).
SN\,2006V deviates strongly from expectations for a \nifs\ powered
BSG star explosion.
\label{fig_a2_06V}
}
\end{figure}

Figure~\ref{fig_a2_06V} compares the $V$-band light curve (with respect to the time of explosion) and the multiepoch spectra of SN\,2006V with model a2 (SN\,1987A and model a4 are also shown in the top panel). The $V$-band light curve of SN\,2006V shows a long rise to a broad maximum at $-17$\,mag, similar to SN\,1987A but brighter by about 1\,mag at all recorded epochs. Its color evolution is however much bluer at all times, and the spectra exhibit much narrower lines than in SN\,1987A \citep{taddia_2pec_12}.

\begin{figure}
   \includegraphics[width=\hsize]{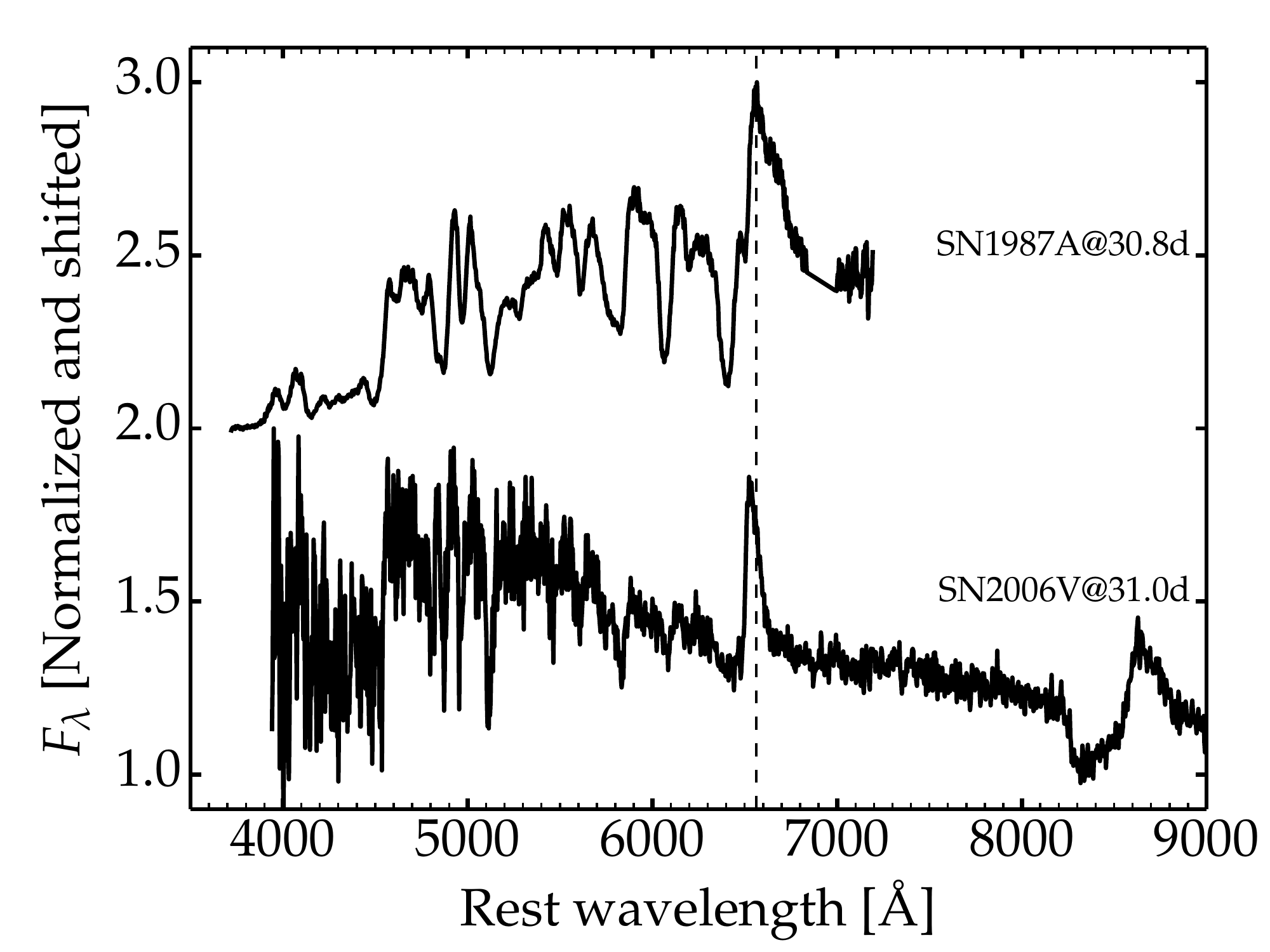}
\caption{Comparison of the observed optical spectra for SN\,1987A and 2006V
at about 30\,d after the inferred time of explosion. The spectra have been corrected
for reddening and redshift,  divided by the maximum flux, and shifted.
This illustrates the stark color contrast between these two SNe at the same post-explosion
phase, as well as the morphological difference in the H$\alpha$ region (with differences
in absorption, strength, width, and overlap with Ba\two\ lines).
\label{fig_87A_06V}
}
\end{figure}

None of our models match the observed properties of SN\,2006V.
Compared to SN\,1987A (see Fig.~\ref{fig_87A_06V} for the comparison
at $\sim$\,30\,d), the narrower spectral lines at all times combined with the
larger brightness break the expectation that more energetic explosions
produce more \nifs\ (the adopted reddening is negligible and increasing it
would increase the discrepancy). The blue spectra with weak signs of blanketing
and narrow lines suggest another power source such as interaction
or a central engine (perhaps combined with a metallicity effect).
Other spectral anomalies point in this direction.
For example, the H$\alpha$ line profile shows a weak absorption trough
(Fig.~\ref{fig_87A_06V}),
reminiscent of the interacting SN\,1998S in which the cold-dense shell formed
during interaction with CSM boosts the H$\alpha$ emission \citep{D16_2n}.
The lack of O\one\,7774\,\AA\ or the Ca\two\ doublet at 7300\,\AA\ is intriguing --
this may be an ionization, a density, or an abundance (e.g., metallicity) effect.
The last spectrum at 101\,d (the exact date depends on the inferred time
of explosion but this spectrum is clearly after $V$-band maximum; \citealt{taddia_2pec_12})
is very odd for a SN that is about to turn nebular because we can see a strong continuum
and no sign of a strengthening line emission.

\subsection{Comparison for SN\,2006au}
\label{sect_spec_06au}

\begin{figure}[h!]
   \includegraphics[width=\hsize]{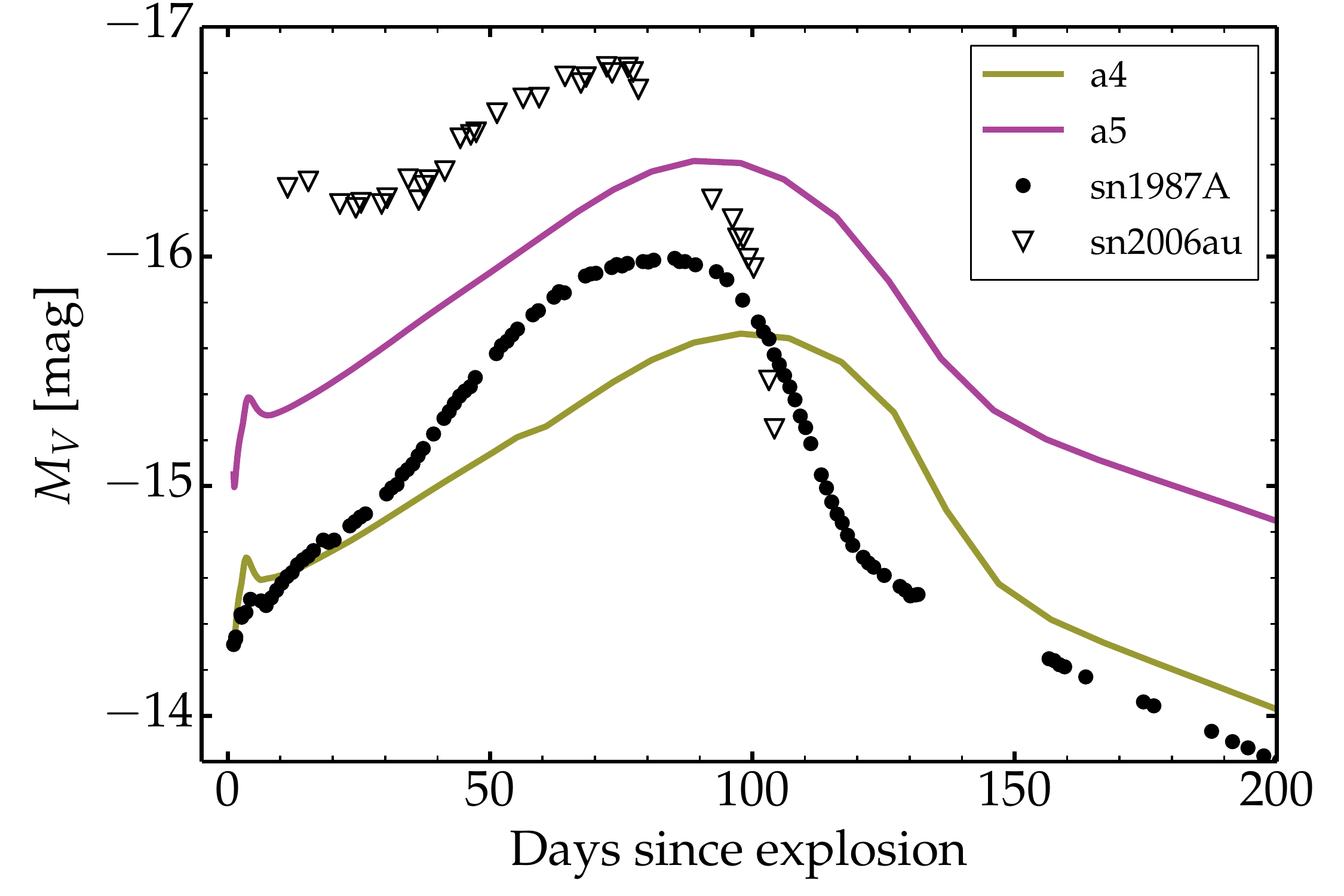}
   \includegraphics[width=\hsize]{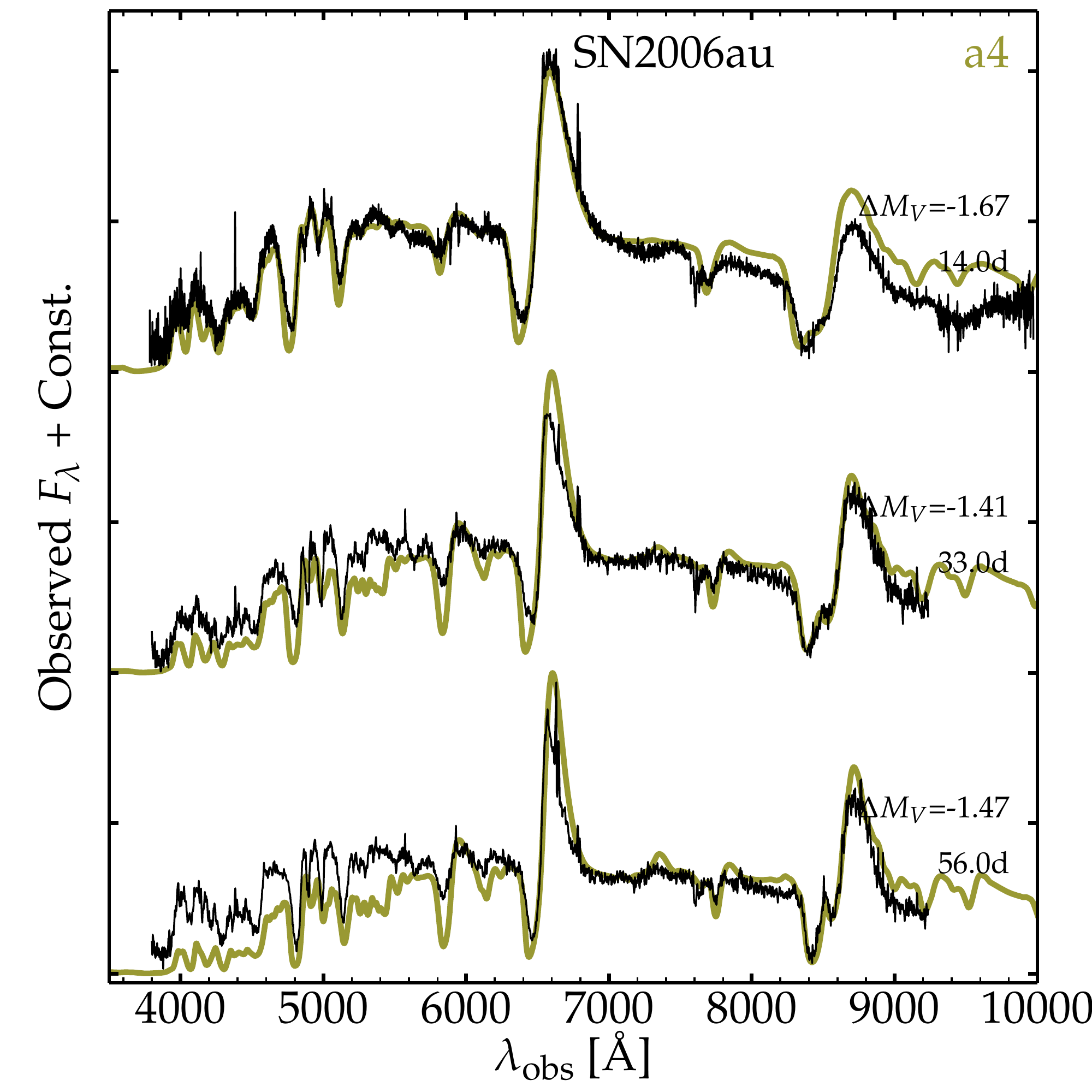}
\caption{Same as Fig.~\ref{fig_a5_00cb}, but now for SN\,2006au
and model a4 (the top panel also includes model a5 and SN\,1987A).
\label{fig_a4_06au}
}
\end{figure}

Figure~\ref{fig_a4_06au} compares the $V$-band light curve and the multiepoch spectra of SN\,2006au with model a4 (SN\,1987A and model a5 are also shown in the top panel). In the $V$ band, SN\,2006au is bright early on, with a magnitude of -16.4\,mag (nearly two magnitudes brighter than SN\,1987A), rises to a maximum of $-16.8$\,mag in about 70\,d (about 15\,d earlier than SN\,1987A), but then precipitously fades to become very faint, which suggests a very low \nifs\ mass. The large brightness and bluer colors early on suggest a bigger progenitor star (say $75-100$\,\rsun) than for SN\,1987A (probably around 50\,\rsun) or our present model \citep{taddia_2pec_12}. Models a4 and a5 have a similar brightening rate as SN\,2006au prior to maximum, but they peak later and are  too bright at nebular times. If the \nifs\ mass is very low in SN\,2006au, it is not clear what powers the SN at maximum. If we adopt a negligible reddening, the color discrepancy disappears and SN\,2006au has a similar brightness to SN\,1987A at maximum.

 The bottom panel of Fig.~\ref{fig_a4_06au} shows that model a4 matches some
 of the spectral properties of SN\,2006au, but the model is too red. The observed
optical color of SN\,2006au is not standard. It has a blue optical color despite
the presence of Na\one\,D, Fe\two\ lines, Ti\two\ lines, which are indicative
that the spectrum formation region is roughly at the H recombination temperature.
The blue color seems to hardly change from 14 to 56\,d and the signs of blanketing
even seem to weaken. There is no spectroscopic information  at nebular times to help
constrain the power source and the characteristics of the inner ejecta.

The properties of SNe\,2006V and 2006au may in part be driven
by a lower metallicity, perhaps much smaller than in the LMC.
The problem is that the optical color should eventually redden
as the photosphere cools and recedes into the more metal-rich parts
of the ejecta. Similarly, a larger radius can yield bluer colors early on,
but this effect tends to ebb as the SN progresses through the recombination
phase. A power source distinct from radioactive decay might resolve this
issue.

  SN\,2004ek shares some similarities with the properties of SNe 2006V and 2006au
\citep{taddia_2pec_16}. It has a similar light curve to SN\,2006au (anomalously
bright early on), similar narrow lines to SN\,2006V (much more narrow than SN\,1987A),
and a peculiar blue color (peculiar blue colors, narrow lines, and relatively high brightness
are also observed in LQS13fn; \citealt{polshaw_lsq13fn_16}).
Its H$\alpha$ profile shows a weak absorption at the recombination
epoch, which is reminiscent of SN\,1998S \citep{D16_2n}. Hence, the peculiar
properties of SN\,2004ek may reflect the influence of an interaction.

\section{Conclusions}
\label{sect_conc}

   We have presented non-LTE time-dependent radiative-transfer simulations for
BSG star explosions. We have used one progenitor model to produce
a variety of explosion models by varying the energy of the thermal bomb
and the prescribed \nifs\ mass (in order to correlate the two). We also investigate
the effect of a different hydrogen to helium abundance ratio, of a different
\nifs\ mixing magnitude, as well as of a different \nifs\ mass.

We use this set
of simulations (drawn from a single progenitor model) to compare to observed
SNe II-pec. The take-home message from SNe II-pec observations is that there
is no prototype for this class. SNe II-pec exhibit a large disparity in optical light curves,
optical color curves, in rise time to maximum, in brightness contrast between maximum
and the nebular phase,  in brightness versus expansion rate.
There is also considerable spectral
diversity. These events show very distinct line profiles.
For example, we see strong Ba\two\ lines
in SN\,1987A but these lines are weak in SN\,2000cb.
Additionally,  there is strong H$\alpha$ absorption in SN\,1987A
while it is weak in SN\,2006V.
So much diversity is hard to explain, and hard to reproduce with just one set of
models. Evidently, explaining Type II-pec SNe requires more than just
a reduction in progenitor radius.
But our model set helps characterize the incongruous features of SNe II-pec.

   Our simulations exhibit a large range in brightness at peak and at nebular times,
which correlates with the \nifs\ mass. All models (as well as SN\,1987A) have
a luminosity at peak about 50\% greater than the contemporaneous absorbed decay
power. For energetic models,  $\gamma$-ray escape starts soon after maximum and
prevents the direct inference of the \nifs\ mass.  In some models (e.g., a5), the change
in IR flux at nebular times compensates for the growing $\gamma$-ray escape to
yield a decay rate consistent with full $\gamma$-ray trapping.

  Our models of BSG star explosions powered by \nifs\ systematically
  enter the recombination phase within at most a week after explosion.
  Subsequently, they evolve at constant color ($U-V$ of about 3\,mag),
  which is also roughly the same for all models in our grid,
  and their luminosity reflects the trajectory of their (optical) photosphere.
  Bolometric maximum occurs when the photospheric radius is maximum.
  Models with a larger explosion energy have a larger photospheric
  radius at a given post-explosion time and consequently a larger
  luminosity. This correlation between radius and luminosity no longer
  holds when clumping is considered \citep{d18_fcl}.
  In that case, clumping can speed up the recession of the photosphere,
  yielding smaller photospheric radii but greater luminosities because
  of the enhanced release of stored radiant energy.

  The brightening rate on the rise to maximum brightness is comparable
  in all our models, except when \nifs\ mixing is tuned.
  Depending on the \nifs\ mass, the rise time to maximum
  lies between 90 and 130\,d. At the earliest times, the luminosity is primarily
  influenced by the explosion energy (we use only one progenitor model, otherwise
  a dependency on radius and envelope density structure would also appear)
  and negligibly by the \nifs\ mass (even with strong mixing).

  The spectral evolution for our model set is very uniform, the only
  diversity arising from line profile widths (and the indirect effect
  of line overlap). The greater the initial explosion energy, the broader
  the line profiles at all times, including at nebular times. At early times,
  \nifs\ plays no role so line broadening is directly related to explosion
  energy. As the SN enters the recombination phase and the photosphere
  recedes in mass space, this correlation can break down if \nifs\ mixing
  is changed. For example, a weaker explosion can exhibit broader lines
  at bolometric maximum if \nifs\ is initially more efficiently mixed
  to large velocities.

  We use this set of models to diagnose the properties
  of observed Type II-pec SNe, in particular to identify, whenever relevant,
  the signatures of asphericity, clumping, chemical inhomogeneities, mixing, or
  a distinct power source from decay heating. We find that in all cases
  our models fail in one way or another to match the observations.
  This in part reflects the scatter in SNe II-pec properties, in particular the fact
  that brighter SNe do not necessarily exhibit broader lines at similar epochs.
  Either the SN luminosity is not powered by \nifs\ decay or the correlation
  between \nifs\ mass and explosion energy has a large scatter in Type II-pec SNe.

  Photometrically (i.e., in the $V$ band), the best match is obtained for model a4 and SN\,2009E, and for a4he and SN\,2009mw, while spectrocopically, the best match is obtained for model a5 and SN\,2000cb. Photometrically, our models tend to rise too slowly and for too long. This is particularly evident for SNe with a short rise time (25\,d for SN\,2000cb; 70\,d for SN2006au, 80\,d for SN\,1987A and SN\,2006V). Although the data is sparse, the best match is for model a4he and SN\,2009mw. The rise time of SN\,2009E is matched by model a4, but this model has much broader spectral lines at all times.

  In some cases, the mismatch between observations and models is
  expected. SN\,2000cb has a 25\,d rise to maximum, which is typical of
  SNIbc. However, SNe Ibc are characterized by (inferred) ejecta masses
  of $3-5$\,\msun\ (see, e.g. \citealt{drout_11_ibc}). In contrast, our model
  has an ejecta mass of 13\,\msun\ (and a standard explosion energy).
  The fast rise might indicate that the explosion was asymmetric
  \citep{utrobin_chugai_00cb_11}, although in our model sample,
  even the highest energy models with the highest \nifs\ mass
  have a slower rise to maximum.

  SNe\,2006V and 2006au are too blue at all times to be explained
  by a standard BSG star explosion. These objects are in tension
  with \nifs\ decay heating. SN\,2006V has narrow lines at all times
  while being the brightest SN II-pec in our sample.
  SN\,2006au is very bright early on, which may suggest a bigger
  radius than for our adopted progenitor star. But its peculiar color
  and abrupt fading from maximum are puzzling.

   Some of these discrepancies suggest that our progenitors or ejecta models
   are inadequate. However, in some cases, the colors of both models
   and observed SNe are in rough agreement and the offset in
   brightness is associated with offsets in photospheric radii
   at the 30\,\% level, which is not so large.

  Spectroscopically, the comparison between our model set
  and our selected sample of SNe II-pec is very informative.
  For SN\,2000cb, model a5 matches the whole spectral evolution
  apart from H$\alpha$  and one epoch (i.e., at 38.9\,d).
  For SN\,1987A, model a4 matches the whole spectral evolution
  except the phase $20-90$\,d. During the rise to maximum,
  the structure of H$\alpha$ is not matched, and H$\beta$ is
  predicted but absent. In SN\,1987A, the photosphere recedes
  much faster than in any of our models, showing line widths
  typical of model a4 initially and of model a2 past 30\,d
  (Fig.~\ref{fig_v_ha}). The large strength of Ba\two\ lines
  in SN\,1987A is unlikely caused by a time-dependent effect since
  our simulation treats this process and the Ba\two\ lines are
  strong only in the model with the lowest explosion energy (a2).
  Furthermore, Ba\two\ lines
  are weak in SN\,2000cb and SN\,2006au. In SN\,2009E,
  Ba\two\ lines are strong and well matched by our model a2.
  Hence, we believe the strength of Ba\two\ lines is controlled
  by ionization primarily. Ejecta with a lower expansion rate
  are denser and produce stronger Ba\two\ lines, as seen
  in low-energy SNe II-P
  \citep{roy_08in_11,lisakov_08bk_17,lisakov_ll2p_18}.
  Clumping can also strengthen the Ba\two\ line strength
  \citep{d18_fcl}.

  We find that the width and strength of H$\alpha$ becomes sensitive
  to \nifs\ decay heating (and the associated non-thermal processes)
  only after $10-20$\,d. Prior to that, the strength of Balmer lines
  is sensitive to non-LTE and time-dependent effects.

The blue colors of SNe 2006V and SN2006au
might in part be caused by the low metallicity of the SNe II-pec
environments. On the modeling side, the mixing we adopt largely erases
this low initial metallicity. Allowing for macroscopic mixing without
microscopic mixing might perhaps help producing bluer colors. If this
chemical segregation and mixing are generic, this notion would not explain
why SNe II-pec like 1987A and 2000cb are much redder than other SNe II-pec
like 2006V or 2006au. For the latter two (including SN\,2004ek;
\citealt{taddia_2pec_16}), a differing powering mechanism
might be responsible for the SN brightness.

 Understanding the diversity of SNe II-pec will require more work. In the future, we will revisit the present work and cover a more extended parameter space,  by including additional progenitor models (mass, radius etc) and a greater variety of explosion characteristics. We also need to consider alternative power sources to decay heating (such as power radiated by the compact remnant and interaction with circumstellar material) as well as investigate the impact of clumping, chemical inhomogeneity, and large-scale asymmetry.

\begin{acknowledgements}

LD thanks ESO-Vitacura for their hospitality.
This work used computing resources of the mesocentre SIGAMM,
hosted by the Observatoire de la C\^ote d'Azur, France.

\end{acknowledgements}

\end{document}